 \documentclass[final,5p,times,twocolumn]{elsarticle}

\usepackage{amssymb}

\usepackage{bookmark}
\usepackage{graphicx}  
\usepackage{dcolumn}   
\usepackage{bm}        
\usepackage{amssymb}   
\usepackage{amsmath}

\begin{document}

\begin{frontmatter}



\title{On the Stability of Fermionic Non-Isothermal Dark Matter Halos}


\author{Ahmad Borzou}
\ead{ahmad\_borzou@baylor.edu}
\address{EUCOS-CASPER, Physics Department, Baylor University, Waco, TX 76798-7316, USA}

\begin{abstract}
The stability of isothermal dark matter halos has been widely studied before. In this paper, we investigate the stability of non-isothermal fermionic dark matter halos.
We show that in the presence of temperature gradient, the force due to the pressure has both inward and outward components. In some regions of halos, the inward force that provides stability is due to the pressure rather than gravity. Moreover, it is shown that higher temperature gradients lead to halos with lower mass and size.
We prove that if the temperature is left as a free positive profile, one can place no phase-space lower bound on the mass of dark matter. 
For halos that are in the low degeneracy classic domain, we derive an analytic expression of their temperature in terms of their mass density and place an upper bound on the mass of dark matter by requiring that temperature is not negative. We then use the Burkert mass profile for the Milky Way to show that if the central temperature of the halo is a few Kelvins, the mass of dark matter cannot exceed a few keV. 
\end{abstract}


\end{frontmatter}


\newcommand{\bq}{\begin{equation}}
\newcommand{\eq}{\end{equation}}
\newcommand{\bqn}{\begin{eqnarray}}
\newcommand{\eqn}{\end{eqnarray}}
\newcommand{\nb}{\nonumber}
\newcommand{\lb}{\label}
\newcommand{\Lnz}{\text{Ln}(z)}
\newcommand{\Lns}{\text{Ln}(s)}
\newcommand{\mnras}{MNRAS}
\newcommand{\apj}{Astrophys. J.}
\newcommand{\prl}{Phys. Rev. Lett.}
\newcommand{\prd}{Phys. Rev. D}
\newcommand{\physrep}{Phys. Rep.}
\newcommand{\aap}{Astronomy and Astrophysics}

\section{Introduction}
A variety of independent observations leave no doubt on the existence of dark matter (DM). These include the early measurements of galaxies velocity dispersion in the Coma cluster \cite{1933AcHPh...6..110Z}, the rotation curves in galaxies \cite{1980ApJ...238..471R}, the recent measurements of the gravitational lensing \cite{2003ARA&A..41..645R, 1998ApJ...498L.107T}, the Bullet cluster \cite{2006ApJ...648L.109C}, the anisotropies in the CMB \cite{2011ApJS..192...19W}, and the large scale structures \cite{2003MNRAS.342..287A}.

Since no candidate particle has been observed in any of the experimental searches \cite{2017PhRvL.119r1301A, 2017PhRvL.118b1303A, PhysRevLett.119.181302, 2018ARNPS..68..429B}, direct measurement of DM properties such as its mass is not possible at present.   
However, galaxies are giant DM laboratories where we can investigate the properties of the constituting particles by relating them to the observable features of DM halos such as their stability. The involved forces are gravity which can be explored by observation and the gradient of the pressure of DM. The latter depends on the equation of state (EOS) of DM and as a result, provides an invaluable possibility to investigate the foundations of these long-sought particles. 

Since the visible mass of galaxies is made of fermions, DM may also have fermionic nature. In this case, the pressure of the halo is described by the Fermi-Dirac statistics. This scenario has been studied before in \cite{2001astro.ph.11366B, 2002PhRvE..65e6123C, 2014MNRAS.442.2717D} for isothermal halos. 
Recently, degenerate models of fermionic DM halo have attracted attention mainly due to their potentials for addressing the core-cusp problem \cite{2013NewA...22...39D, 2015JCAP...01..002D, 2017MNRAS.467.1515R, 2018MNRAS.475.5385D}.

In the Fermi-Dirac statistic, the pressure is a function of the temperature and the fugacity--or equivalently the chemical potential. 
Therefore, the force due to the pressure has in general two components. 
The first force of the pressure is due to the derivative of the fugacity, which is investigated in references above within the isothermal halos. The second force of the pressure is due to the temperature gradient. As we will see in this paper, one of the two forces of the pressure can be inward, adding to the pulling gravitational force and deepening the potential well. 

In general, fermions can be compressed in arbitrarily small volumes despite the restriction on their phase-space. Black holes are the living examples of compressed fermionic systems. Even though the phase-space of fermions is limited, there is an unlimited momentum space available to them. By reducing their configuration volume, they will occupy higher momentum states to meet the Pauli exclusion principle. It is only the balance of the forces due to their pressure and gravity that determines the size of fermionic DM halos. As we will see in this paper, the inward force of the pressure can be the dominant pulling force in a vast region of the halo that confronts the pushing force of the pressure and maintains stability.

In this paper, we discuss that a constant temperature across DM halos is not validated through observations. Moreover, even for collision-less DM models, there are a variety of heat generation mechanisms that can lead to at least slight deviations from isothermal models. 
We review the gravitational and non-gravitational frictions and gravitational contraction as the main sources of heat generation, and radiation and convection as the means of heat transfer in DM halos. The frictional effects may play crucial roles in satellite galaxies that move with relatively high speeds through their host galaxies. Also, the gravitational contraction mechanism of heat generation is more important in these compact halos than in larger dilute halos. Therefore, as we will show in this paper, it is possible that the compactness of satellite galaxies is due to their different temperature profiles. 

We show that if DM distribution in the halos is Maxwellian, i.e. fermions are entirely at low degeneracy level, the temperature profile can be expressed analytically in terms of the mass profile. By requiring non-negative temperatures, we put an upper bound on the mass of DM. Also, we analytically prove that the temperature profile is irrelevant if DM halo is entirely in the infinite degeneracy level. 

To study non-isothremal fermionic DM halos, we derive the most general hydrostatic equilibrium equation for a spherical Fermi-Dirac system. Computer software is developed to solve the field equations numerically. We validate the software by reproducing the known solutions of different degeneracy levels.
Since the software is meant to be general, rather than assuming a specific DM and galactic model to derive the temperature as a function of the radius, we reserve the assumption for directly inserting a temperature profile into the software. 

We study non-isothermal models with a generic temperature profile of the form $T = T_0\left(1+\left(\frac{r}{r_0}\right)^2\right)^{-1}$, in a range of low central temperature of $T_0 \leq 1$ Kelvin. We show that the isothermal solution $r_0=\infty$ leads to the largest halo size and mass. By increasing the temperature gradient, the mass and the size of the halo decrease. 
We present solutions that are more compressed than their corresponding infinitely degenerate halos.

The current lower-bound on the mass of DM is derived using highly compressed infinitely degenerate isothermal solutions \cite{2018MNRAS.475.5385D}. However, in the presence of temperature gradient, more compressed halos are possible; allowing for the possibility of lighter DM masses. 
We discuss that since (i) the pressure and the mass density of Fermi-Dirac systems are functions of two independent profiles of temperature and fugacity, and (ii) there is only one stability equation in terms of the mass density and pressure, any arbitrarily light dark matter mass can explain any observationally supported mass profile if the temperature is left as a free positive profile. 
Therefore, phase-space lower bound on the mass of DM requires modeling the non-universal temperature profile of DM halos.

This paper is organized as follows. 
In section \ref{Sec:Fermi-Dirac stat}, the statistics of Fermi-Dirac systems and their stability criteria are reviewed. 
In section \ref{Sec:StabilityMostGeneral}, we derive the most general stability equation for halos made of fermions and present computer software to solve it. 
In the same section, we study a class of solutions whose temperatures decrease with distance from the center. 
Also, we place an upper bound on the mass of DM using the stability of low degenerate halos. 
 In section \ref{Sec:PhaseSpaceMassBounds}, we discuss the phase-space lower bounds on the mass of DM and show the relevance of temperature profiles for them. 
A conclusion will be drawn in section \ref{Sec:conclusion}.

\section{An overview of the Fermi-Dirac statistics and non-isothermal halo stability}
\lb{Sec:Fermi-Dirac stat}
The visible matter in galaxies is made of fermions. It is quite likely that DM also has the same nature. 
In general, both temperature and mass density are functions of the radial distance from the center of halos. Therefore, the degeneracy of fermionic DM can be at the opposite extreme levels at two locations of one same halo. Therefore, the most general EOS of the fermionic matter and its corresponding stability equation is of interest. In this section, we review the statistics of non-isothermal fermionic systems and the corresponding stability equation. We also briefly discuss the sources that can establish temperature gradients in DM halos. 

\subsection{EOS of Fermi-Dirac statistics}
Since the mass density, and the pressure are local quantities measured in the free-falling frame at a given distance $r$, it is easiest to derive the EOS of fermionic DM from its energy-momentum tensor in the same frame.
Calculations in this frame are particularly advantageous if the corrections to the EOS due to possible interactions between DM are of interest.

The energy density and the pressure operators of the system are the temporal and the spatial components of the energy-momentum tensor associated with the Lagrangian of a non-interactive Dirac spinor field
\bqn
\lb{Eq:EnergyDensityPressure}
&&\rho_{\text{energy}} = \frac{1}{V}\int d^3x T^0_0, \nb\\
&&P = \frac{1}{3}\int d^3x \left( T^1_1 + T^2_2 + T^3_3 \right),
\eqn
where the integral is over microscopic scales and $V$ is a local volume.

In a free-falling frame, the gravitational effects are absent, and the metric is Minkowskian. Working in this frame, we can insert the free field expansion of the fermionic field into equation above 
to derive the energy density and pressure operators. 
The pressure, the energy density, and the mass density
are the ensemble averages of the corresponding quantum operators, which using the non-relativistic approximation, read
\bqn
\lb{Eq:PressureDensity}
&&P = \frac{2(kT)^{\frac{5}{2}}}{\alpha^3}f_{\frac{5}{2}}(z),\nb\\
&&\rho_{\text{energy}} = \frac{3}{2}P,\nb\\
&&\rho = \frac{2m(kT)^{\frac{3}{2}}}{\alpha^3}f_{\frac{3}{2}}(z),
\eqn
where $z$ is the fugacity, $\beta = \frac{1}{kT}$, and $\alpha \equiv \frac{h}{\sqrt{2\pi m}}$ in terms of the Planck constant. 
Also, the Fermi-Dirac integrals are defined as 
\bqn
f_{\nu}(z)=\frac{1}{\Gamma(\nu)}\int_0^\infty\frac{x^{\nu-1}dx}{z^{-1}e^x+1},
\eqn
with $\Gamma(\nu)$ being the gamma function.
Moreover, it is useful to know that the derivative of the Fermi-Dirac integrals read 
\bqn
\frac{d f_{\nu}(z)}{dr} = f_{\nu-1}(z)\frac{d\left(\Lnz\right)}{dr}.
\eqn

The most general EOS for non-interacting fermionic DM can be easily read from Eq. \eqref{Eq:PressureDensity}
\bqn
\lb{Eq:EOS}
P = \frac{kT}{m}\rho h(z),
\eqn
where 
\bqn
h(z) \equiv \frac{f_{\frac{5}{2}}(z)}{f_{\frac{3}{2}}(z)}.
\eqn

When the degeneracy is high, i.e. $z \gg 1$, we can use the Sommerfeld approximation
\bqn
\lb{Eq:SommerfeldPerturb}
f_{\nu}(z) \simeq \frac{\left(\Lnz\right)^{\nu}}{\Gamma(\nu+1)}\left(
1+\frac{\pi^2\nu(\nu-1)}{6} \left(\Lnz\right)^{-2} + \cdots
\right).
\eqn
In the case of very low degeneracy level, we can use the following approximation
\bqn
\lim_{z\rightarrow 0} f_{\nu}(z) \simeq z,
\eqn
when $h(z)\simeq 1$, and we recover the EOS of the classic ideal gas.

\subsection{Conservation of energy-momentum, stability equation, and temperature profile}
\lb{Sec:Conservation}
In a stable solution, the net force on an arbitrarily small volume of mass is zero at any point of the halo. The hydrostatic equilibrium equation can be systematically derived using the conservation of the energy-momentum tensor together with Newton's field equation and reads
\bqn
\lb{Eq:HydroStat1}
\frac{dP}{dr} \simeq -G \rho\frac{ M(r)}{r^2},
\eqn
where $G$ is Newton's constant, and $M(r)$ is the mass enclosed within the distance r.

If DM annihilation and creation are not significant, the conservation of the energy-momentum also implies that
\bqn
\lb{Eq:MassConservation}
M(r) = \int_0^r 4\pi r'^2 \rho(r') dr'.
\eqn 
This and Eq. \eqref{Eq:HydroStat1} can be combined into a second order differential equation
\bqn
\lb{Eq:HydroStat2}
\frac{1}{r^2}\frac{d}{dr}\left(\frac{r^2}{\rho}\frac{dP}{dr}\right) \simeq -4\pi G \rho,
\eqn
where the pressure and density are local, i.e., measured in the free falling frame at distance $r$. Also, the equation is valid regardless of the statistics of DM.

Finally, we would like to mention that the equilibrium equation above can be equivalently derived in a frame attached to the center of halo, rather than to a free-falling observer. 
In this frame, the energy of DM particles receives an extra gravitational potential energy which should be absorbed by the chemical potential, i.e. the latter is also different in the two frames. 
The disadvantages of this frame are that the equations contain both astronomical and microscopic lengths, the equations are not manifestly covariant anymore, and in the case of strong gravitational effects, cumbersome calculations are needed to arrive at Tolman-Oppenheimer-Volkoff stability equation.

\begin{figure}
\centering 
\includegraphics[width=1\linewidth]{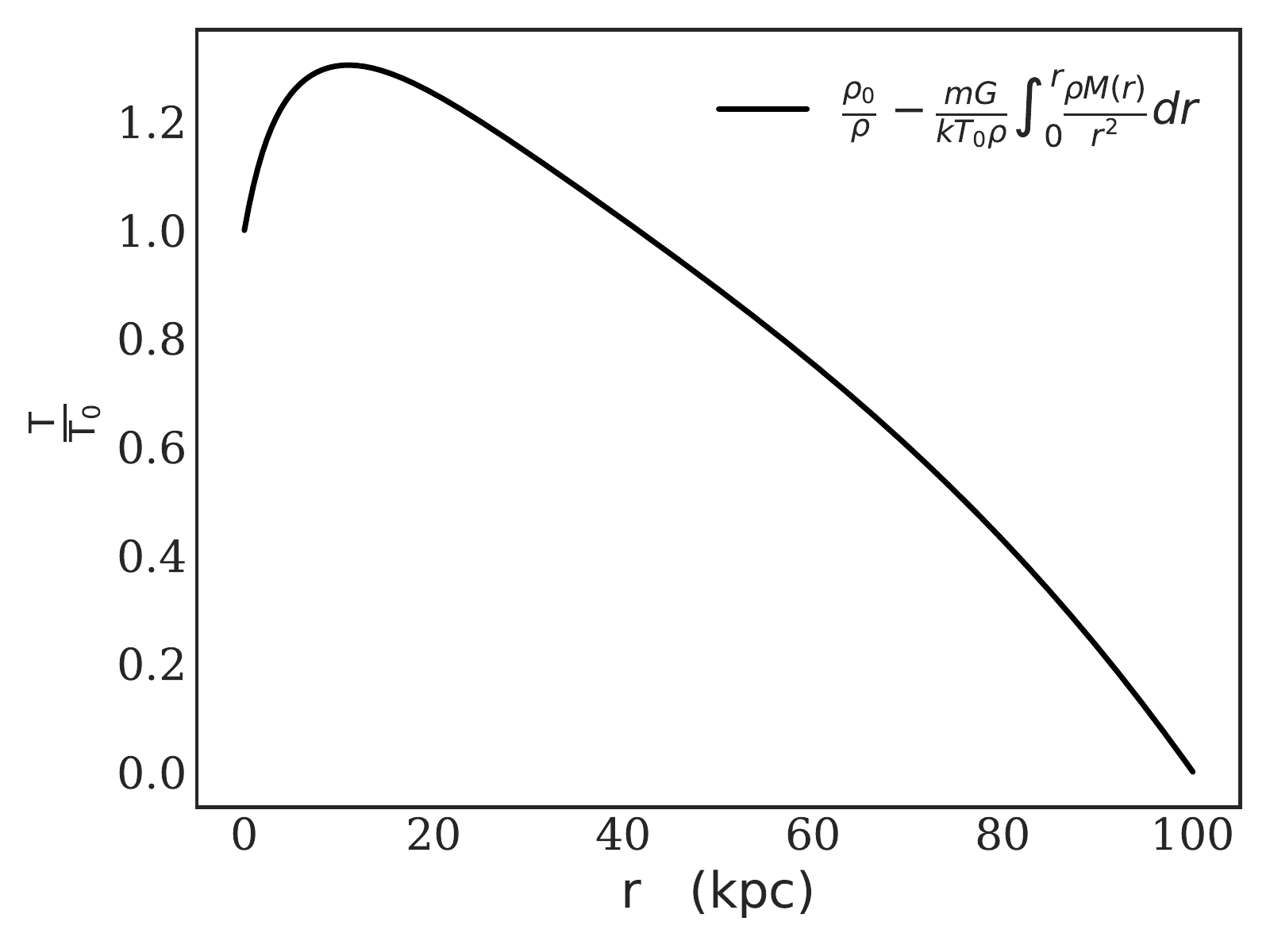}
\caption{\lb{Fig:PredictedTemperatureForBurkertMassProfile}
Predicted temperature profile for a Burkert mass density with $r_0=10$~(kpc) and $\rho_0=2.7\times 10^{-21}$~(kg$\cdot$m$^{-3}$) and $m/T_0=542$~(eV$\cdot$K$^{-1}$).}
\end{figure}

\subsection{Sources of heat generation and heat transfer}
\lb{Sec:TempGrad}
Since the stability equation depends also on the gradient of temperature, it is important to model the temperature profile of DM halos. 
In this section, we show that unlike the mass density with the equation for hydrostatic equilibrium, the temperature profile does not have a unique differential equation. It not only depends on the model of DM but also is a function of galactic properties such as relative speeds, the position of other galaxies, and the mass densities. We specifically discuss the means of heat transfer like radiation and convection and the origins of the heat, such as gravitational contraction, tidal forces, and friction.

\paragraph{Heat generation}
Any phenomenon through which the macroscopic kinetic energy of galaxies is transformed, due to the conservation law, to the microscopic kinetic energy of the constituting particles, falls under the category of friction. The drag forces felt by compact dwarf galaxies that are moving through their hosts may set up considerable temperature gradients. Such forces usually depend on the relative speed, the mass densities, and other environmental properties. In collision-less models of DM, the Chandrasekhar friction is the most significant drag force. In interacting models of DM, a variety of frictional forces are possible. For QED-like interactions, the drag force can be approximated by the familiar form of the friction in the fluid dynamics.

Conversion of the gravitational potential energy to the kinetic energy of DM is another means for heating galaxies. The well-known Kelvin-Helmholtz mechanism belongs to this category.   

Tidal stirring is another means of heating a halo. Most of the satellite galaxies orbiting their host experience  a tidal force due to the difference between the host's gravitational force at the front and rear edges of the satellite.

\paragraph{Heat transfer}
In collision-less DM models, only gravitational sources are available for radiation. The Kelvin-Helmholtz mechanism is an example. 
If DM interacts through a long-range force, in SIDM category, for example, DM can cool down through dark radiations, for instance see \cite{ 2014PhRvD..90d3524B, 2014MNRAS.445L..31B, 2018PhRvD..97l3004B}. The Eddington equation for temperature gradient due to radiation reads \cite{1967aits.book.....C}
\bqn
\frac{dT}{dr} = - \kappa \frac{\rho L}{T^3 r^2},
\eqn
where $\kappa$ is a constant and $L$ is the luminosity of the radiation at distance r from the center.

Convection is the most efficient means of heat transfer and is available to both collision-less and interacting DM models. 
It can be shown that the condition for the convection in a Fermi system reads
\bqn
\lb{Eq:ConvectionStability_0}
\frac{dT}{dr} \leq \frac{2T}{5P}\frac{dP}{dr} -\frac{kT^2\rho}{mP}\frac{dh}{dr},
\eqn
where
the right hand side is the adiabatic temperature gradient for a general fermionic gas. 
If the heat transfer is fast enough and the galaxy under study is convective, one can take the equality in this equation to determine the temperature gradient. 

%

\section{Stability solutions of fermionic DM halo}
\lb{Sec:StabilityMostGeneral}
In this section, we would like to study the stability of DM halo under any degeneracy level. We first discuss the extreme degeneracy scenarios analytically and then present computer software for the sake of investigating the degeneracy regions that cannot be explored analytically.

\subsection{Temperature profile of a DM halo with Maxwellian distribution}
\lb{Sec:TemperatureProfileMaxwell}
In a wide range of models, DM particles are assumed to have Maxwellian distribution. Therefore, their halos are described by the EOS of a classic ideal gas which reads
\bqn
P = \frac{kT_0}{m} y \rho, 
\eqn
where naught refers to the values at the center, $m$ is the mass of DM, $k$ is the Boltzmann constant, and $T\equiv T_0 y$ is the temperature at arbitrary distance from the center.
This EOS combined with Eq.~\eqref{Eq:HydroStat1} implies that 
\bqn
\frac{d\left(y\rho\right)}{dr} = -\frac{m G}{k T_0} \frac{\rho M(r)}{r^2}. 
\eqn

If the temperature is constant everywhere in the halo, $y\rho=\rho$, and a direct substitution confirms the stability of the well-known isothermal solution
\bqn
\lb{Eq:ClassoIsoSolution}
\rho = \frac{k T_0}{2\pi m G}\frac{1}{r^2}. 
\eqn
This solution is singular at the center, i.e. does not satisfy the initial conditions, and also is in contradiction with observations of mass density at around the center. 

The temperature profile can be derived through the integration of differential equation above
\bqn
\lb{Eq:AnalyticTemperatureClassic}
y(r) = \frac{\rho_0}{\rho} - \frac{m G}{k T_0}\frac{1}{\rho}\int_0^r \frac{\rho M(r')}{r^{'2}}dr',
\eqn
where $y_0=1$ is used. 

Since temperature cannot be negative at any distance from the center, we can derive an upper bound on the mass of dark matter
\bqn
\lb{Eq:UpperMassClassic}
m \leq \frac{k T_0 \rho_0}{G}\frac{1}{\text{Max}\left(\int_0^r \frac{\rho M(r')}{r^{'2}}dr'\right)}.
\eqn
Since possible dark matter interactions are not strong, $T_0$ is not expected to be high, and we should be able to find a fair maximum value for it. Therefore, the upper mass for dark matter can be set by observing the mass profile of different halo types.  

As an example, in \cite{2013JCAP...07..016N}, the preferred mass profile for the halo of the Milky Way is found to be the Burkert profile
\bqn
\rho = \frac{\rho_0}{\left(1 + \frac{r}{r_0}\right)\left(1+(\frac{r}{r_0})^2\right)},
\eqn
with $\rho_0 \simeq 2.7\times 10^{-21}$~(kg$\cdot$m$^{-3}$) and $r_0 \simeq 10$~(kpc). Therefore, $\int_0^r \frac{\rho M(r')}{r^{'2}}dr'$ rise with the distance until reaches a flat plateau of $\sim 0.58$ kg$^2\cdot$m$^{-4}$--hence the maximum value, at around 30 (kpc). 
Inserting these into Eq.~\eqref{Eq:UpperMassClassic}, we find that 
\bqn
\frac{m}{T_0} < 542 \left(\frac{\text{eV}}{\text{K}}\right).
\eqn
If $T_0$ is around a few Kelvin, the mass of DM cannot exceed a few keV. This can be converted to a lower bound for the dispersion velocity at the center of the halo
\bqn
\sigma_0 \equiv \sqrt{\frac{kT_0}{m}} > 120 \left(\frac{\text{km}}{\text{s}}\right).
\eqn 
The predicted temperature profile of the Milky Way with the largest $\frac{m}{T_0}$ is plotted in figure~\ref{Fig:PredictedTemperatureForBurkertMassProfile}. 


Finally, we would like to emphasize the role of visible matter that has been neglected in the preceding analysis. In the central regions of most of the galaxies, 
the visible matter can have a higher mass density than DM, see for example the simulations in \cite{2015MNRAS.451.1247S,2016MNRAS.461L..11H}. Therefore, the gravitational force due to visible matter in the interior of galaxies is greater than the force due to DM. Consequently, $M(r)$ in Eq.~\eqref{Eq:HydroStat1} should be replaced by $M(r)+M^*(r)$ where asterisk refers to the visible matter. This means that the force of pressure of DM should now confront the force of gravity due to both DM and visible matter. Therefore, the upper bound on the mass of DM in Eq.~\eqref{Eq:UpperMassClassic} should be modified by $M(r')\rightarrow M(r')+M^*(r')$ in the denominator. Due to the extra gravitational force, DM should be even lighter than when the visible matter is neglected.

\subsection{Temperature profile of rotating halos}
\begin{figure}
\centering
\includegraphics[width=1\columnwidth]{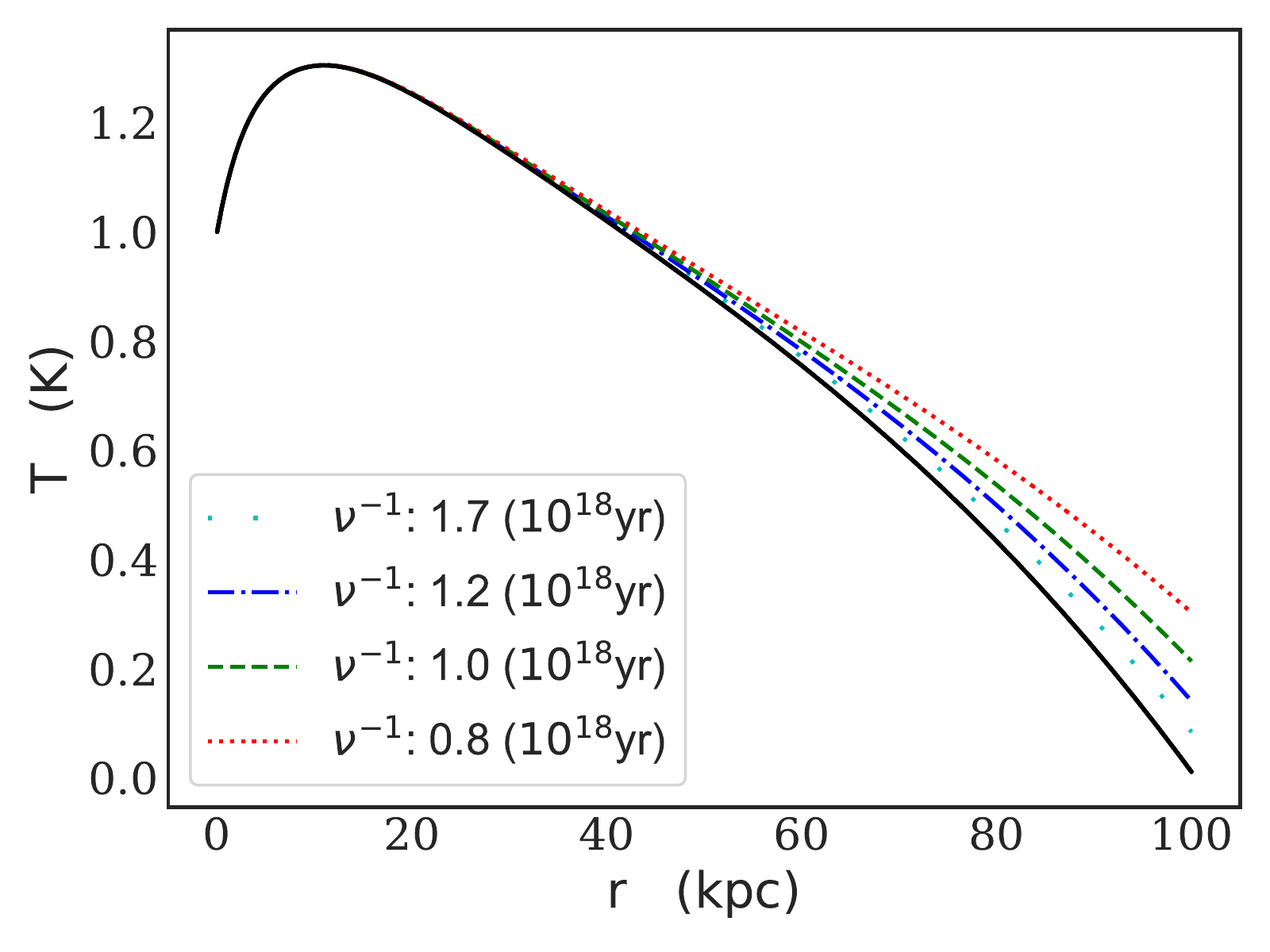}
\caption{Temperature profiles along the radial coordinates $x_1$ and $x_3$ in a cylindrical system where the latter is the height and the former refers to the polar plane. Halos are rotating along the $x_3$ axis with constant angular velocity $2\pi\nu$. The non-solid lines show the profiles along the $x_1$ coordinate while all four halos have the same profiles along $x_3$ shown with overlaying solid lines.  DM mass and the rest of the halo properties are chosen to be the same as the halo presented in figure~\ref{Fig:PredictedTemperatureForBurkertMassProfile}. The solid lines in the two plots match. 
The equatorial edges of the rotating spherical halo should be heated up to maintain stability.  
\lb{Fig:TRotating}}
\end{figure}
In general, the stability equation for a halo rotating with constant angular velocity $\omega$  reads
\bqn
\vec{\nabla}P = \rho\left(-\vec{\nabla}\Phi+\omega^2  x_1 \hat{x}_1\right),
\eqn
where $x_1$ is the radial coordinate in the plane normal to the rotation axis in a cylindrical coordinate system and hat refers to the unit vector. The last term on the right hand side is the centrifugal force while the rest of the terms are the same as before. 

In an isothermal halo, this stability equation can be met only if the halo takes a non-spherical shape of an ellipsoid. However, the shape of a non-isothermal halo does not necessarily depend on its rotation. The reason is that (i) the temperature of a halo is not necessarily a function of its shape and can be changed by halo's environment, and (ii) temperature gradient adds an extra component to the force due to the pressure.

As an example, in this section, we present the temperature profiles of four perfectly spherical but rotating halos with rotation periods of 0.8, 1, 1.2, and 1.7~$\left(\times 10^{18}~\text{yr}\right)$. For the sake of comparison, we choose their mass density, DM mass, and the rest of the parameters to be the same as the halo presented in figure~\ref{Fig:PredictedTemperatureForBurkertMassProfile}. 
The temperature profiles that enforce the stability are shown in figure~\ref{Fig:TRotating}. The very bottom overlying solid lines are the profiles along the axis of rotation $x_3$ and as expected are the same as the temperature profile of the non-rotating halo. The non-solid lines are temperature profiles along the radial coordinate $x_1$ on the $x_3=0$ plane. Although the mass density is still spherical, the temperature profile has deviated from spherical symmetry. The equatorial edges of these rotating halos are heated up to confront the centrifugal force while the central regions are left intact. 

In summary, halo rotation breaks the spherical symmetry of the stability equation, which can be accommodated by breaking the spherical symmetry of mass density, or temperature profile, or both. The complication is that the temperature is also a function of the environment and therefore is not fixed by mass density. This leads to a dramatically high number of possible stable solutions that can be ruled out only after the temperature profile is modeled. 
For mainly this reason, we only consider non-rotating spherical halos in the rest of this paper and leave the more general stability solutions for later.

\subsection{Temperature profile of a fermionic DM halo at full degeneracy level}
\lb{Sec:TemperatureProfileFullDeg}
At the full degeneracy level, $z\simeq \infty$, and using the Sommerfeld approximation, the Fermi-Dirac integrals read  
\bqn
\lim_{z\rightarrow \infty} f_{\nu}(z) = \frac{\left(\Lnz\right)^{\nu}}{\Gamma(\nu+1)}. 
\eqn
Therefore, the ratio of the Fermi-Dirac integrals can be written in terms of the mass density
\bqn
h(z) = \frac{\Gamma\left(\frac{5}{2}\right)}{\Gamma\left(\frac{7}{2}\right)}
\left(\frac{\Gamma\left(\frac{5}{2}\right)\alpha^3}{2m}\right)^{\frac{2}{3}}
\frac{\rho^{\frac{2}{3}}}{kT}.
\eqn 
The EOS in Eq.~\eqref{Eq:EOS} reads
\bqn
\lb{Eq:FullDegEOS}
P = \frac{\Gamma\left(\frac{5}{2}\right)}{\Gamma\left(\frac{7}{2}\right)}
\left(\frac{\Gamma\left(\frac{5}{2}\right)\alpha^3}{2m^{\frac{5}{2}}}\right)^{\frac{2}{3}}\rho^{\frac{5}{3}},
\eqn
which unlike the EOS of the classical ideal gas, has no temperature dependence. Consequently, the temperature profile becomes irrelevant in the full degeneracy limit, and the stability Eq.~\eqref{Eq:HydroStat2} leads to the well-known Lane-Emden equation with numerically known solutions.

So far, we have studied the solutions for when every location of DM halo is at extremely low or full degeneracy levels. Since DM halo is denser and hotter at the center and less dense and colder at the edge, the degeneracy level of DM can vary with the distance from the center. 

\subsection{Software}
Study of the most general scenario calls for numerically solving the most general stability equation because (i) the dependence of pressure on the mass density and temperature, both of which functions of $r$, takes a complicated nature in the partial degeneracy level, and (ii) the known solutions of specific degeneracy levels still depend on the initial conditions that are not known when transiting from one solution to the other. 

Inserting the pressure and mass density from Eq. \eqref{Eq:PressureDensity} into Eq. \eqref{Eq:HydroStat2}, the most general hydrostatic equilibrium equation for non-interacting fermions reads
\bqn
\lb{Eq:HydroStat_Dimensionless}
\frac{1}{\xi^2}\frac{d}{d\xi}\left(
\frac{5}{2} \xi^2 h(s) \frac{dy}{d\xi}+
\xi^2y\frac{d\left(\Lns\right)}{d\xi}
\right)
= - y^{\frac{3}{2}}f_{\frac{3}{2}}(s),
\eqn
where the dimensionless variables are defined as 
$\xi \equiv \sqrt{\frac{8\pi G m^2 (kT_0)^{\frac{1}{2}}}{\alpha^3}}r$, and
$s   \equiv \frac{z}{z_0}$.
The boundary conditions are $s_0=y_0=1$ which are implied by the definitions of $s$ and $y$ and $\frac{ds}{d\xi}|_{\xi=0}=\frac{dy}{d\xi}|_{\xi=0}=0$
which can be understood from Eq.~\eqref{Eq:HydroStat1} knowing that $M(r)$ approaches zero faster than $r^2$ when we move toward the center.

The numerical solutions of differential equation \eqref{Eq:HydroStat_Dimensionless} are sought in terms of $y$ and $\Lns$, instead of $s$. The latter choice is because the fugacity can take computationally infinite values.  
We have written computer code in Python \cite{ComputerSoftwareBorzou}, to study the solutions to the most general stability Eq. \eqref{Eq:HydroStat_Dimensionless} for fermionic non-interacting halos. 
The mass of DM, the central values for the number density and the temperature, and the temperature profile in terms of $\xi$ are taken from the user as inputs. 
The latter is to preserve the generality of the software and allows linking it to data-driven optimization methods for the phenomenological investigation of halo temperatures. 
The reason for this approach is that the equation for temperature gradient not only depends on the DM model but also is a function of the environmental variables of a specific galaxy.

The software starts from the center of the galaxy and uses the input values to compute the rest of the parameters like $\alpha$ and $z_0$. Next, the numerical step size is determined using the Richardson extrapolation at the center. Depending on the magnitude of the error, the step size $(\Delta r)$ can be between one-hundredth of a parsec (pc) to one parsec. 

The software moves toward the edge until the density reaches one-thousandth of its value at the center. 
To reduce the numerical errors to the order of $(\Delta r)^4$ while preserving the computation speed, at each iteration step we use the Verlet method to find $\Lns$ and its first derivative using $\frac{d^2\left(\Lns\right)}{d\xi^2}$ from equation \eqref{Eq:HydroStat_Dimensionless}.

At any iteration, the software determines if the system is in the high or partial or non-degenerate regime. It then uses the appropriate approximations to calculate the Fermi-Dirac integrals. In the case of partial degeneracy where no asymptotic behavior is known, the software uses the optimization methods to learn the fugacity from the Fermi-Dirac integrals at that point. The code will be in its slowest mode when encountering the partial degeneracy level because, unlike the extreme degeneracy levels, given the mass density and temperature, it is not trivial to eliminate $f_{\frac{5}{2}}(z)$ of the pressure in Eq.~\eqref{Eq:PressureDensity} in terms of $f_{\frac{3}{2}}(z)$ of the mass density in the same equation. 

The returned result will be in the form of a set of five plots showing the mass density in units of the critical mass density $\rho_{\text{c}}\simeq 9\times 10^{-27} \text{kg}\cdot \text{m}^{-3}$, the temperature, natural logarithm of the fugacity, the mass of the galaxy, and the chemical potential in the free-falling frame $\mu \equiv kT \Lnz$. The software also reports the dynamical time, the total gravitational potential energy, and the total kinetic energy of the halo. All of the non-isothermal solutions reported in this paper have total kinetic energy equal to approximately half of their gravitation potential energy and as a result, are in the Virial equilibrium. 
Since the full Fermi-Dirac EOS is exclusively used in the software, the transition between degeneracy levels is smooth and the Pauli exclusion principle is always sustained. 
Every reported solution is re-derived using step sizes of one-tenth of the original step size to assure their stability against the global accumulation of the numerical errors.  

\subsection{Software validation}
\lb{Sec:SoftWareValidation}
\begin{figure*}
\centering 
\includegraphics[width=1.\linewidth]{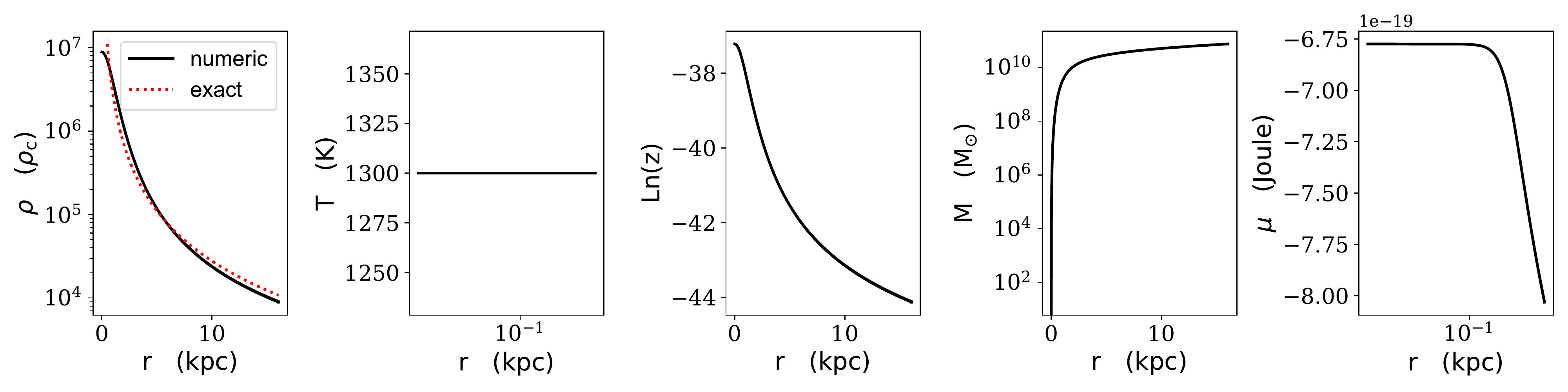}
\caption{\lb{Fig:ClassicIso}A classic isothermal galaxy with DM mass of 1 MeV and temperature of 1300 Kelvin. The red dashed line indicates the exact solution which suffers from a singularity at the center. The initial conditions are set at the distance of 1 pc from the center to avoid the singularity.}
\end{figure*}
Since our code can be used to study any possible solution for non-interacting fermionic DM halo, and for validation purposes, in this section, we re-derive a set of known stability solutions. 

We start with an isothermal model with DM mass of 1 MeV at the temperature of 1300 Kelvin, and a mass density of $8\times 10^{-20}\text{kg}\cdot \text{m}^{-3}$ at 1 parsec from the center. The initial parameters are chosen such that the system is in the classical regime where the hydrostatic equation has an exact solution given in equation \eqref{Eq:ClassoIsoSolution}. 
Figure~\ref{Fig:ClassicIso} shows the numeric and exact solutions for the given parameters. 
The mass density is the first plot from the left, is in fair agreement with the exact solution even though the latter does not satisfy the initial conditions due to its singularity at $r=0$. The solution also contradicts observations due to its cuspy nature at the center. The second plot from the left is the temperature profile showing that it is assumed constant over the halo. The third plot is the logarithm of the fugacity indicating that $z \ll 1$ at all times and confirms the classical nature of the solution. The fourth plot is the mass of the halo. It does not reach a flat plateau indicating another problem of this solution. The last plot is the chemical potential in the free falling frame. Since the temperature is constant, it is proportional to the logarithm of the fugacity in the third plot. However, the x-axis is in the logarithm scale providing the lost information at around the center.

\begin{figure*}[t]
\centering 
\includegraphics[width=\linewidth]{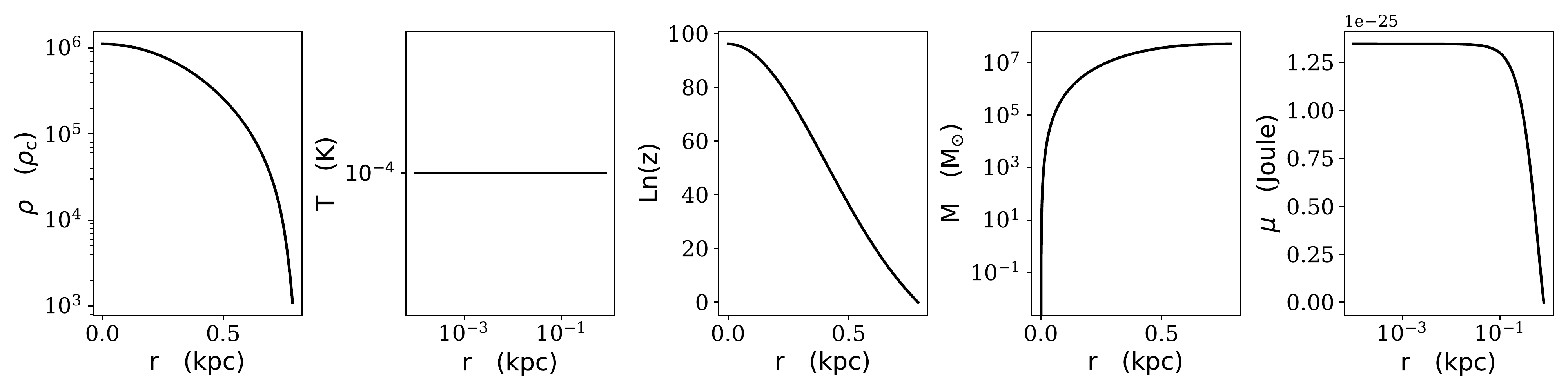}
\caption{\lb{Fig:FullDegDwarfIso}A highly degenerate isothermal DM halo made of 200 eV particles.}
\end{figure*}
To validate the software in the opposite of the spectrum, we reproduce a fully degenerate DM halo presented in \cite{2015JCAP...01..002D} where a lower DM mass limit of 200 eV is derived using the observed line of sight dispersion velocity of dwarf galaxies. We set the DM mass to 200 eV at the temperature of $10^{-4}$ Kelvin, and mass density of $10^{-20}\text{kg}\cdot \text{m}^{-3}$ at 1 parsec from the center. The profile of the system is shown in figure~\ref{Fig:FullDegDwarfIso} where the mass density and total mass of the system are in agreement with those reported in \cite{2015JCAP...01..002D}. 
Since the temperature is not exactly zero, the logarithm of the fugacity is decreasing instead of being infinite as in \cite{2015JCAP...01..002D}. Nevertheless, as far as this logarithm is large enough, the full degeneracy regime is approximately valid and the results are stable. 

\begin{figure*}[t]
\includegraphics[width=\linewidth]{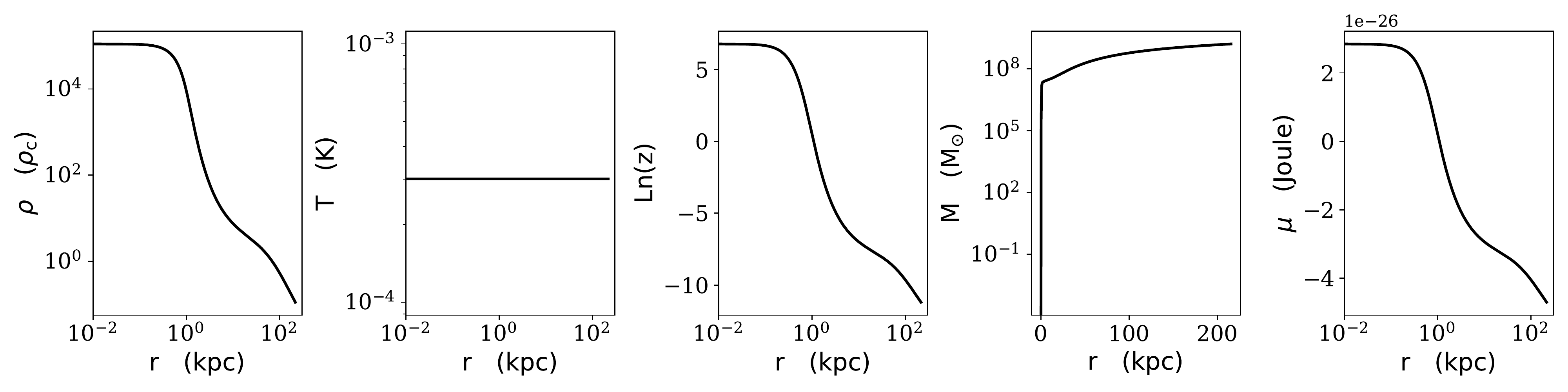}
\caption{Density profile of a halo with DM mass of 200 eV that is highly degenerate at the center, partially degenerate in the middle and non-degenerate at the edge.\lb{fig:doublePlatuea}}
\end{figure*}
In the two solutions above, the entire DM halo was either non-degenerate or highly degenerate. 
We now reproduce the double plateau isothermal solution in \cite{2001astro.ph.11366B,2014MNRAS.442.2717D} where DM halo is highly degenerate at the center, partially degenerate in the middle, and non-degenerate close to the edge. To achieve such solution, we choose the mass of DM to be 200 eV, the density at the center to be $\rho_0 = 10^{-21}\text{kg}\cdot \text{m}^{-3}$, and the temperature at the center to be $T_0 = 0.0003$ Kelvin. The solution is depicted in figure~\ref{fig:doublePlatuea}. It should be noted that, unlike in the previous two solutions, both of the axes of the mass density are transformed into the logarithmic scales to reproduce the looks of the corresponding solutions in the references. Also, to capture the second plateau, we did not terminate the code until the density became $10^{-6}$, instead of $10^{-3}$, times the density at the center. 

So far, we have reproduced the known isothermal solutions. Since non-isothermal solutions are not well investigated before, such comparison is not possible in that domain. However, we still have the analytic evaluations of sections \ref{Sec:TemperatureProfileMaxwell} and \ref{Sec:TemperatureProfileFullDeg} that, as we will discuss later, can validate a subset of non-isothermal solutions. For the rest of the solutions, their numerical accuracy is validated by first running the code in the normal mode and re-running it with intervals of one-tenth of the interval in the normal mode. We only report the solutions if the two trials lead to the same numerical solutions. We would like to mention that for extremely steep temperature gradients, the latter validation may fail, these results are not presented in this paper. It is always possible to reduce the intervals of the numerical method until a valid solution is reached at the cost of slower computations. However, we postpone the study of such domains until more advanced numerical methods are implemented. 

\begin{figure}
\centering
\includegraphics[width=.9\columnwidth]{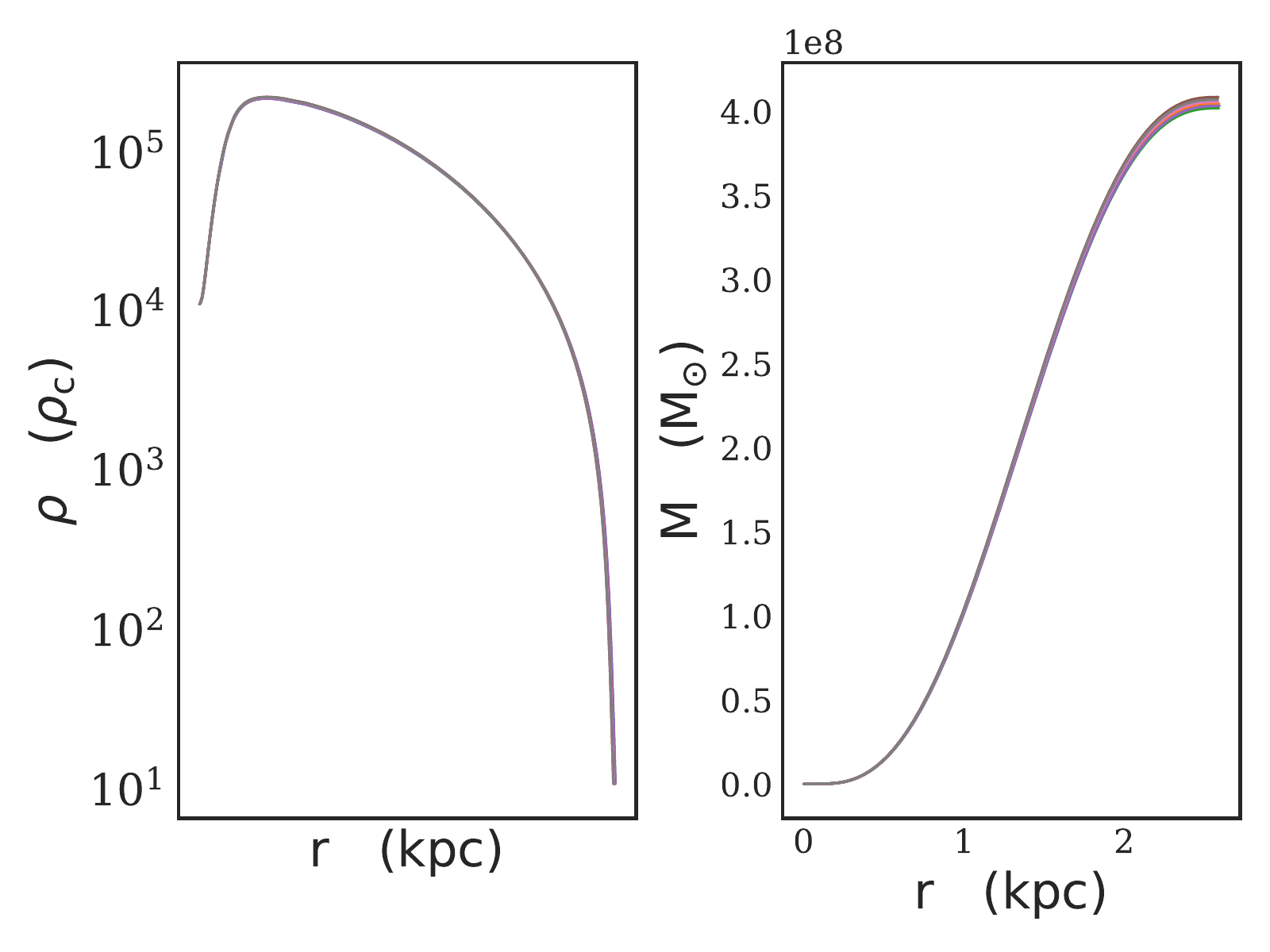}
\caption{Eight perturbed solutions for a halo made of DM masses $m=100$ and $m=100.1$~eV with central temperatures of $T_0=0.1$ and $T_0=0.101$ Kelvin, temperature profile of $y = \left(1 + b \xi^2\right)^{-1}$ with $b=50$ and $b=50.1$ , and central mass density of $\rho_0=10^{-22}$ (kg$\cdot$m$^{-3}$). The eight overlaying curves indicate  the stability of software's results against small changes in the input values.  \lb{Fig:SmallChanges}}
\end{figure}
Finally, the reported solutions are stable against small changes in the input values. An example of such study is shown in figure~\ref{Fig:SmallChanges} where a solution with central temperature of $T_0=0.1$ Kelvin, temperature profile of $y = \left(1 + 50 \xi^2\right)^{-1}$ , and central mass density of $\rho_0=10^{-22}$ (kg$\cdot$m$^{-3}$) is plotted together with solutions of seven small perturbations around the input values as described in the caption. Since all eight curves are overlaying, the perturbation effects are negligible.

\subsection{Non-isothermal DM halos with $y = \left(1 + b \xi^2\right)^{-1}$ as the temperature profile}
\lb{Sec:NonIsoHalo1}
\begin{figure*}
\includegraphics[width=\textwidth]{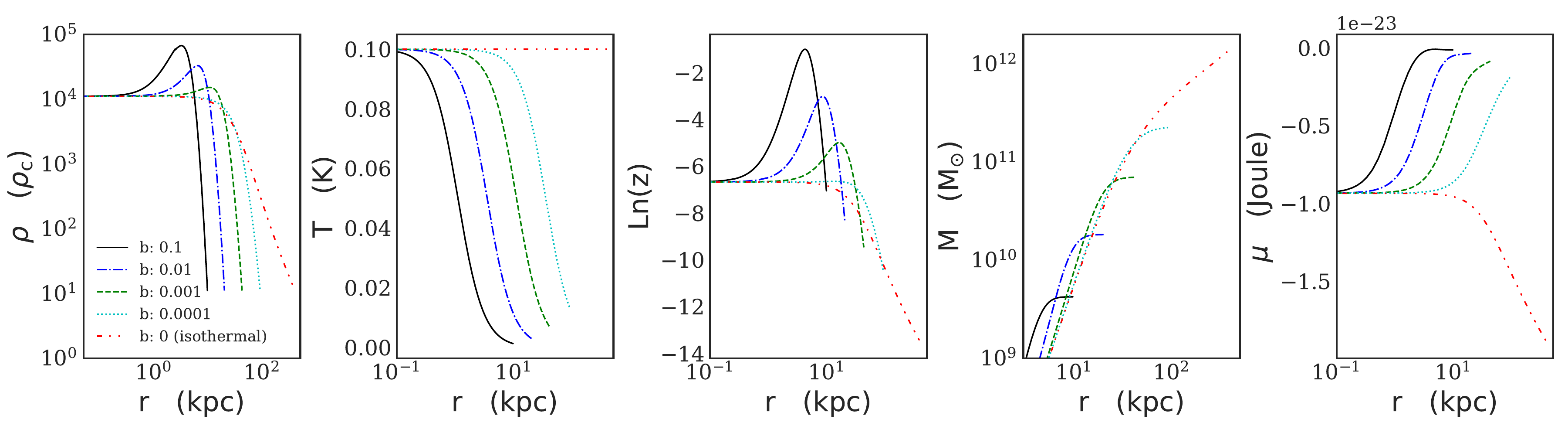}
\includegraphics[width=\textwidth]{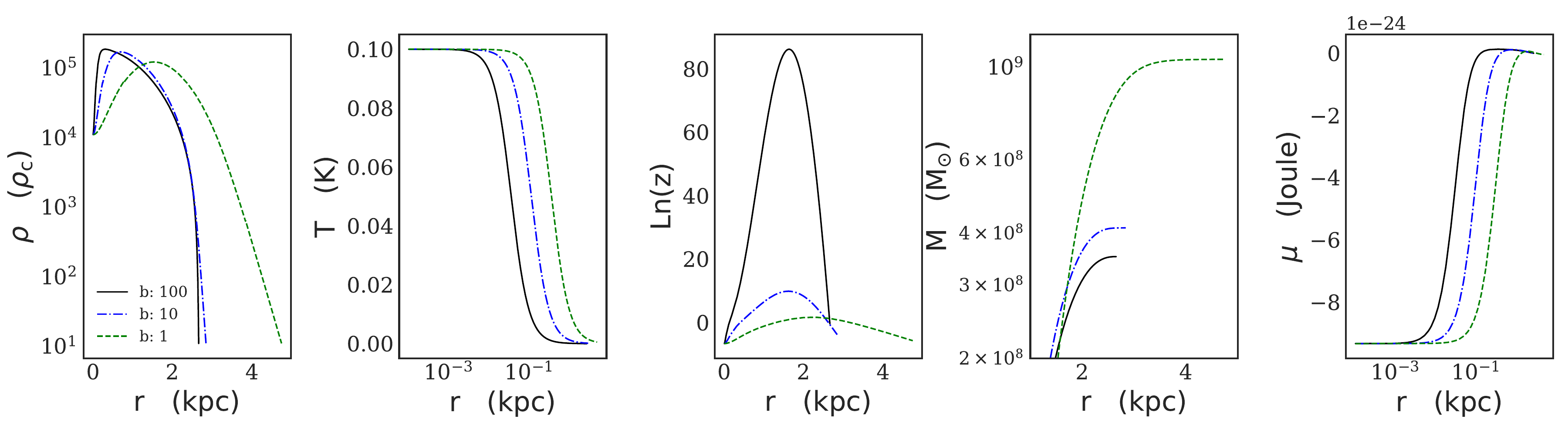}
\caption{Numerical solutions for non-isothermal temperature profile of the form $y = \left(1 + b \xi^2\right)^{-1}$ with $m=100$ eV, $\rho_0=10^{-22}$ (kg$\cdot$m$^{-3}$), and $T_0=0.1$ (Kelvin) and different non-isothermality parameters $b$.\lb{Fig:AllCurves1}}
\end{figure*}
A physically acceptable temperature profile does not take negative values and also does not increase monotonically with the distance from the center. It also must satisfy the initial conditions $y_0=1$ and $\frac{dy}{d\xi}\big|_0=0$. We confirm that 
\bqn
\lb{Eq:AssumedTemperature1}
y = \frac{1}{\left(1 + b \xi^2\right)},& b \geq 0,
\eqn
with $b$ controlling the level of non-isothermality, satisfies all these requirements. 

\begin{figure}
\centering
\includegraphics[width=0.9\columnwidth]{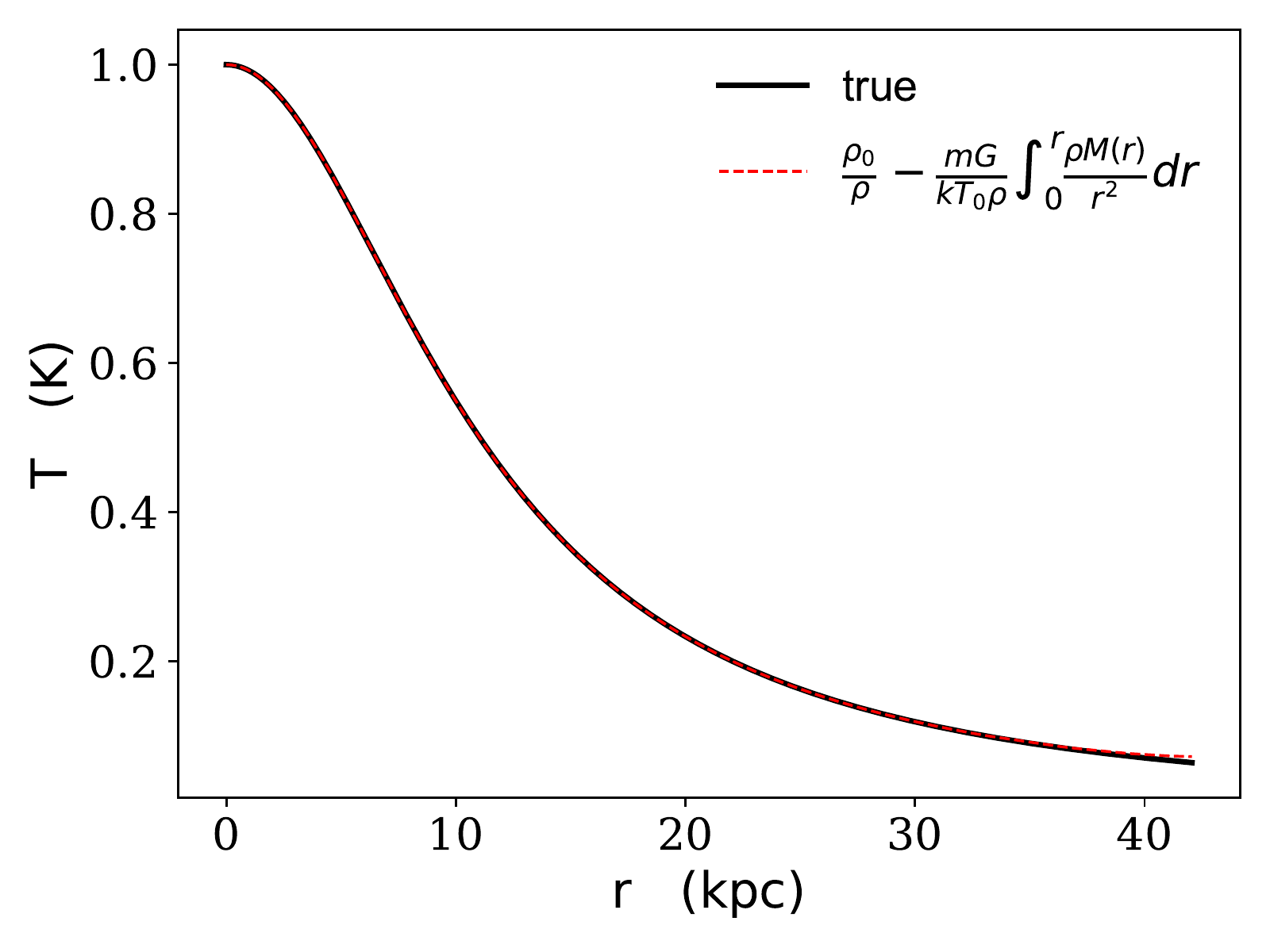}
\caption{The temperature profile $y = \left(1 + b \xi^2\right)^{-1}$ with $b=0.001$ as the true curve and the predicted profile using Eq.~\eqref{Eq:AnalyticTemperatureClassic} and assuming that the mass density is given in figure~\ref{Fig:AllCurves1}.\lb{fig:AnalyticConfirm_b_0.001}}
\end{figure}  
First, we study non-isothermal effects by choosing DM mass of 100 eV--lowest possible DM mass if halos are isothermal \cite{2018MNRAS.475.5385D}, with central temperature of $T_0=0.1$ Kelvin, and central mass density of $\rho_0=10^{-22}$ (kg$\cdot$m$^{-3}$) with different values of $b$. The results are shown in figure~\ref{Fig:AllCurves1}. 

Before analyzing the outputs of the software, we would like to discuss their validity.  
The curve of $b=0$ is the classic isothermal solution. We validated the software in this domain in the previous section in figure~\ref{Fig:ClassicIso}. From the third subplot, we can see that the solutions of $b=0.0001$ and $b=0.001$ are in the domain of low degeneracy with $z \ll 1$ at any distance from the center. Therefore, the analytic formula in Eq.~\eqref{Eq:AnalyticTemperatureClassic} should explain the temperature profile. We validate the latter by inserting the numerical mass density from the software into Eq.~\eqref{Eq:AnalyticTemperatureClassic} and comparing the predicted temperature profile with the assumed one. This comparison is presented in figure~\ref{fig:AnalyticConfirm_b_0.001} where the analytic temperature profile of software's mass profile is predicted to be exactly the same as the assumed one. 
The other curves with higher $b$ values enter the partial degeneracy level at some distance from the center. Their accuracy is validated by re-running the code with reduced distance intervals as discussed in the previous section. 

The very left subplot of figure~\ref{Fig:AllCurves1} shows the predicted mass density. For small temperature gradients with $b \leq 0.1$ (top panel), the mass density starts to increase at some large distance from the center and then drops to zero. Even the classical analytically validated solution for $b=0.001$ possesses a clear overdensity at around 10 (kpc).  For larger temperature gradients with $b\geq 10$ (bottom panel), the mass profile has the shape of a doughnut with a ``hole'' at the center with a negligible radius. This dilute region is located where the visible matter and the central black hole are and may not be detectable. The observable central density will be the peak of the curve and is higher than what we have inserted in the software.  

The total masses of the halos in figure~\ref{Fig:AllCurves1} reach a flat plateau in non-isothermal solutions while continues to infinity in the isothermal case. Therefore, non-isothermal halos are more consistent with expectations. 

The very right subplot in figure~\ref{Fig:AllCurves1} shows that 
even the lowest degree of non-isothermality leads to an increasing chemical potential in terms of $r$ as opposed to the decreasing chemical potential of the isothermal solution. 
 
Figure~\ref{Fig:AllCurves1} also indicates that by increasing the temperature gradient, the radius, and the mass of halo decrease allowing for more compressed solutions. As we will see below, the two solutions with $b\geq 10$ (bottom panel) are more compressed than their corresponding fully degenerate halos--as the most compact scenarios in isothermal cases.

\begin{figure}
\centering
\includegraphics[width=0.9\columnwidth]{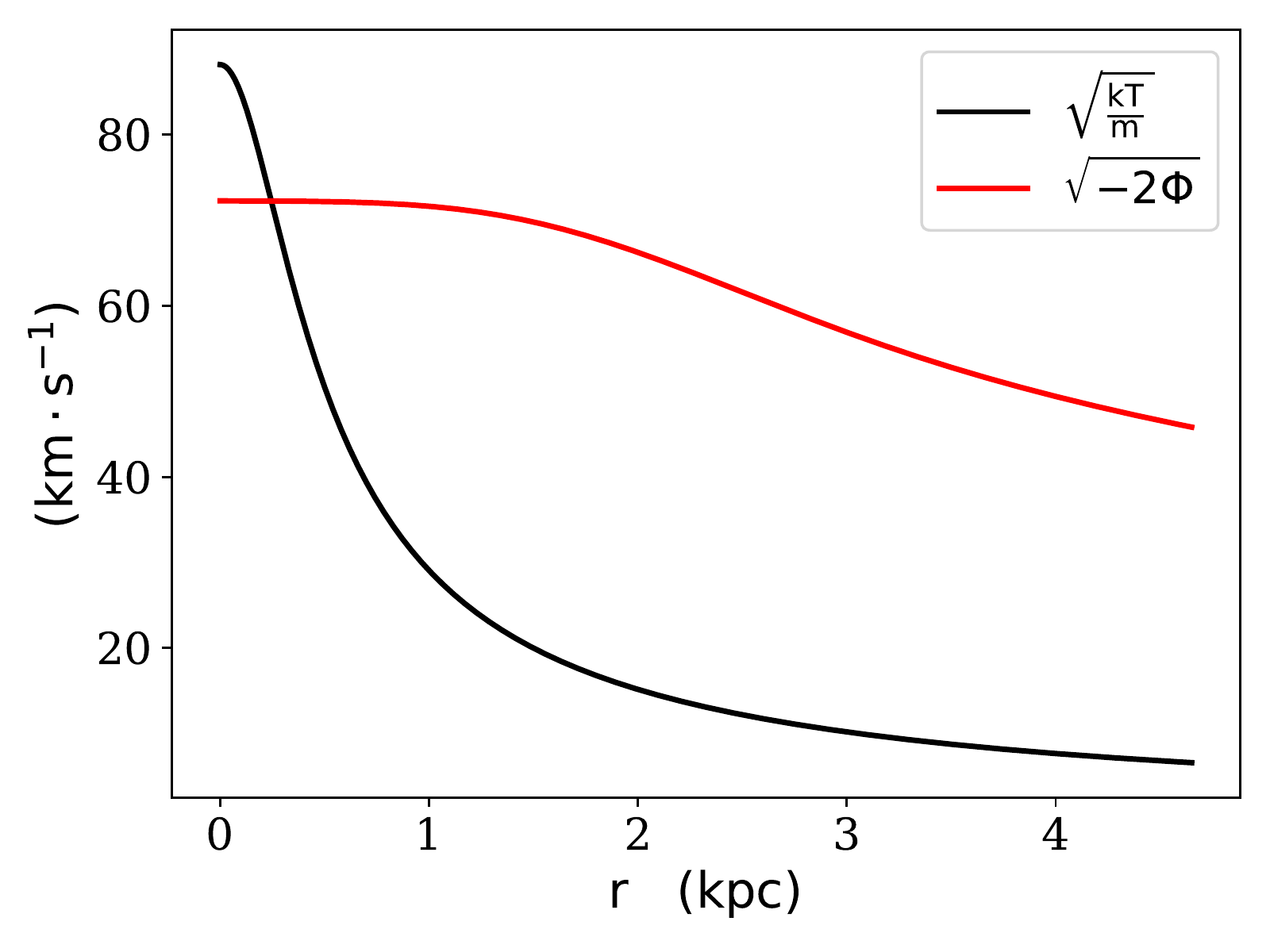}
\caption{\lb{Fig:escapeVelocity}
The escape velocity $\sqrt{-2\Phi}$ and $\sqrt{\frac{kT}{m}}$ corresponding to $b=1$ of figure~\ref{Fig:AllCurves1}. DM particles at the center have higher kinetic energy than the gravitational potential energy. Near the edge, however, their kinetic energy falls toward zero due to the decrease in their temperature. The dynamical time of the system is $1.8\times 10^8$ years. If an imaginary mass bubble moves from the center toward the edge, this extended period assures us that it will lose its kinetic energy to the colder surrounding and cannot escape the system by the time it reached the edge.}
\end{figure}
We would like to mention that, unlike the solutions with constant temperature, in non-isothermal scenarios, the dispersion velocity of DM particles decreases with the distance from the center as can be seen in figure~\ref{Fig:escapeVelocity}. Although the kinetic energy of DM particles is higher than the gravitational potential energy at the center, the kinetic energy at the edge is negligible and the dispersion velocity is much less than the escape velocity.

\begin{figure*}
\includegraphics[width=\textwidth]{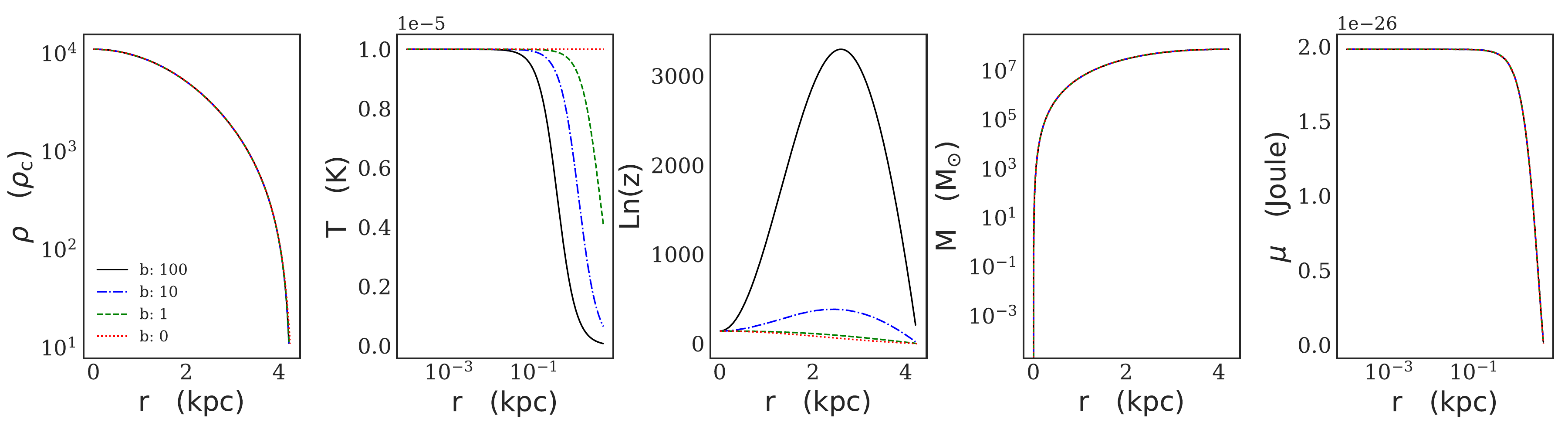}
\caption{Numerical solutions for non-isothermal temperature profile of the form $y = \left(1 + b \xi^2\right)^{-1}$ with $m=100$ eV, $\rho_0=10^{-22}$ (kg$\cdot$m$^{-3}$), and $T_0=10^{-5}$ (Kelvin) and different levels of non-isothermality level $b$. This plot indicates the irrelevance of the temperature profiles since all of the mass density solutions overlay.\lb{Fig:AllCurves2}}
\end{figure*}
To explore the temperature profile in the high degeneracy level scenarios, we lower the central temperature to $T_0=10^{-5}$ (Kelvin) but keep the DM mass and central mass density to be $m=100$ eV and $\rho_0=10^{-22}$ (kg$\cdot$m$^{-3}$) respectively. From section~\ref{Sec:TemperatureProfileFullDeg}, we expect no dependence on the temperature profile and consequently on the value of $b$. The results for different $b$ settings are shown in figure~\ref{Fig:AllCurves2}, and confirm the validity of our software in this domain. 

\begin{figure*}
\centering
\includegraphics[width=0.4\textwidth]{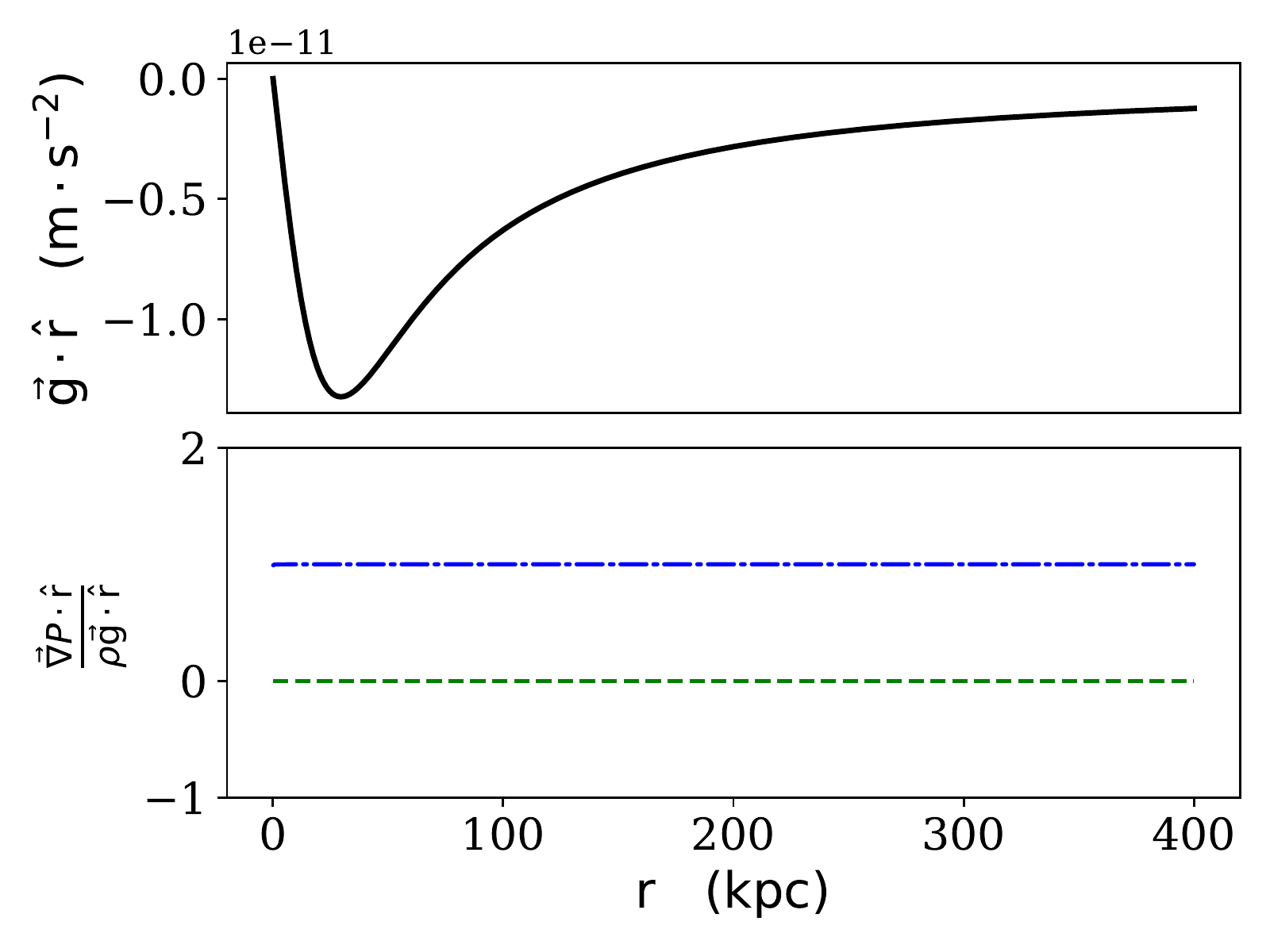}
\includegraphics[width=0.4\textwidth]{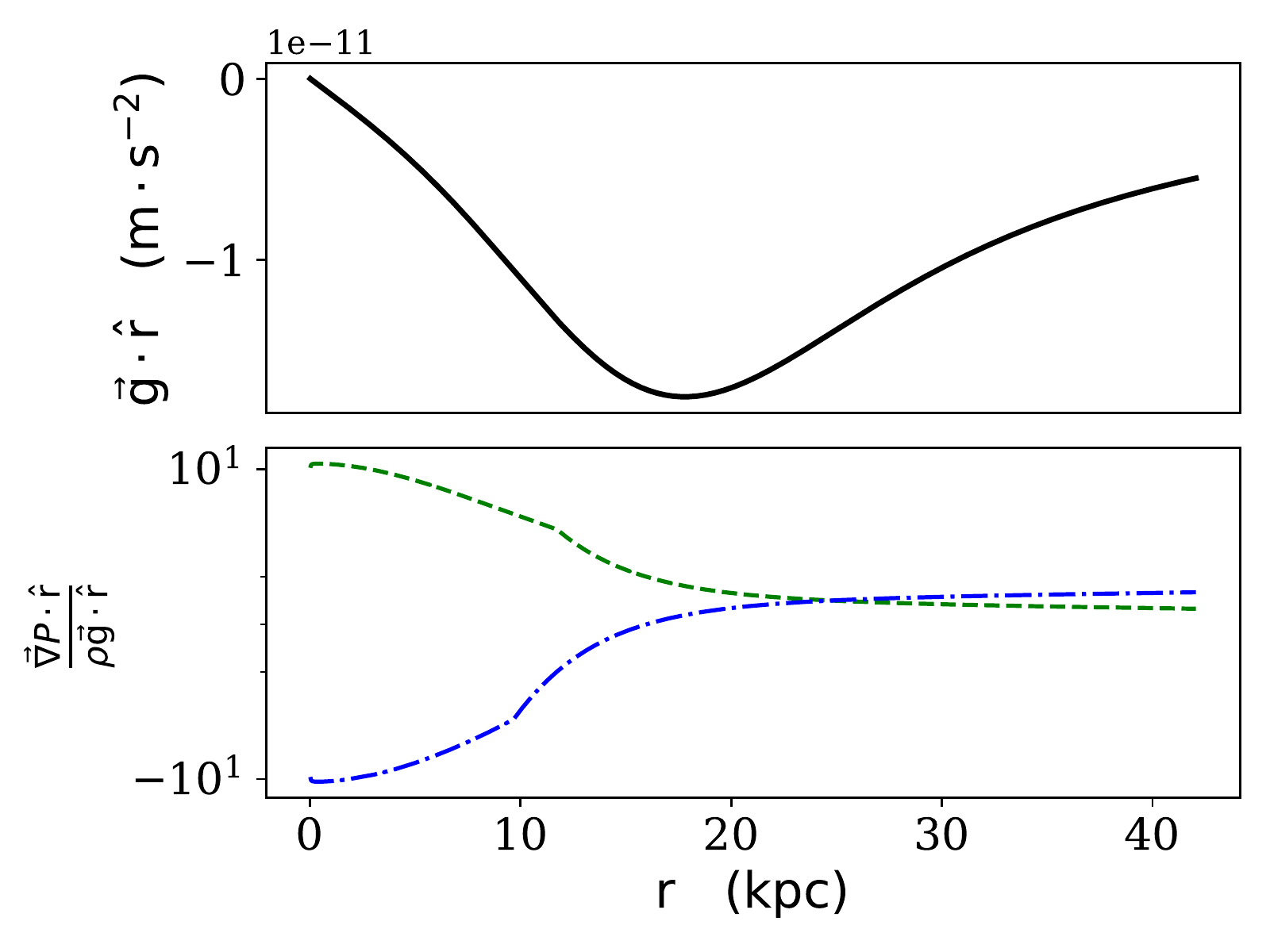}
\includegraphics[width=0.4\textwidth]{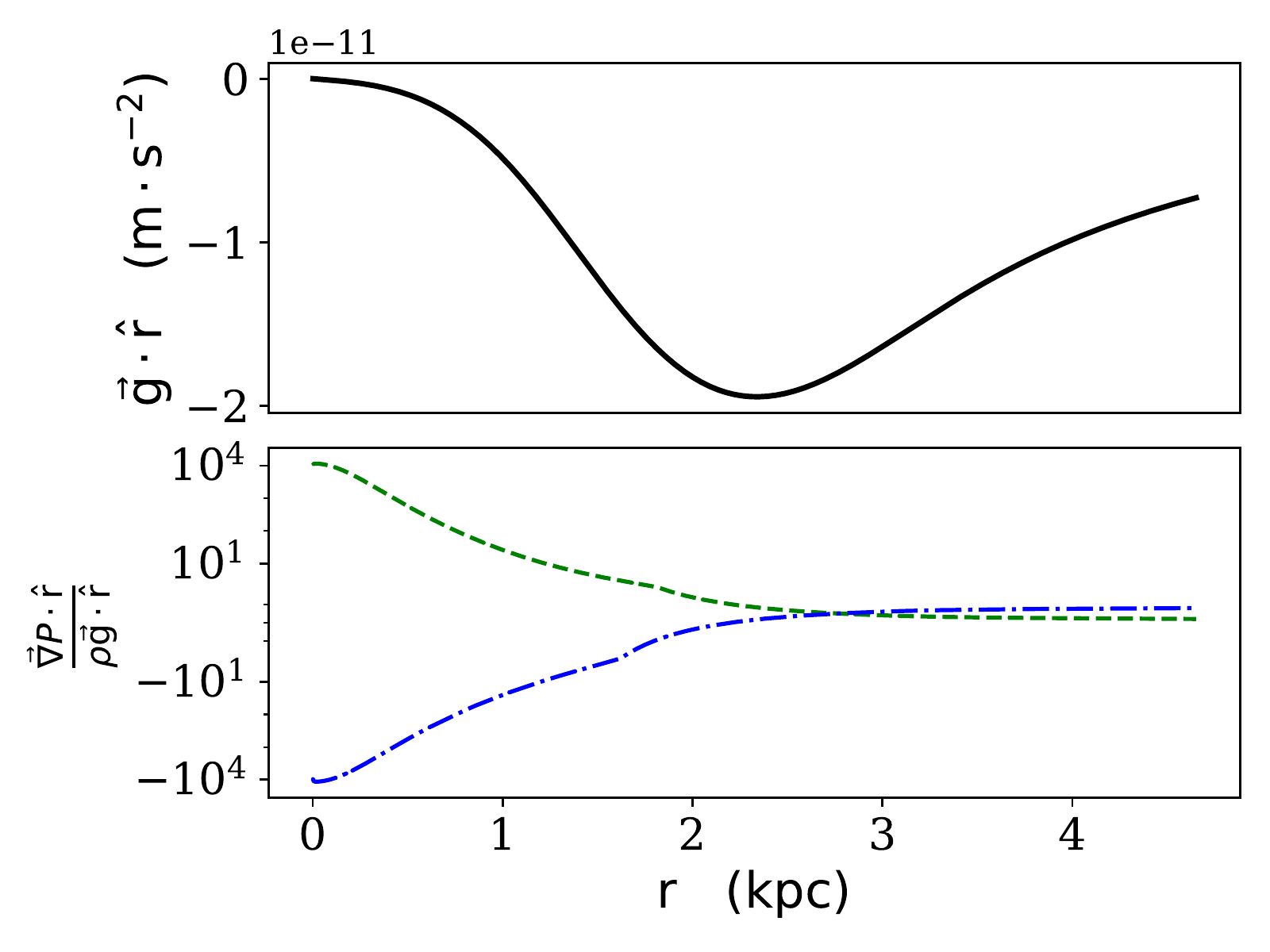}
\includegraphics[width=0.4\textwidth]{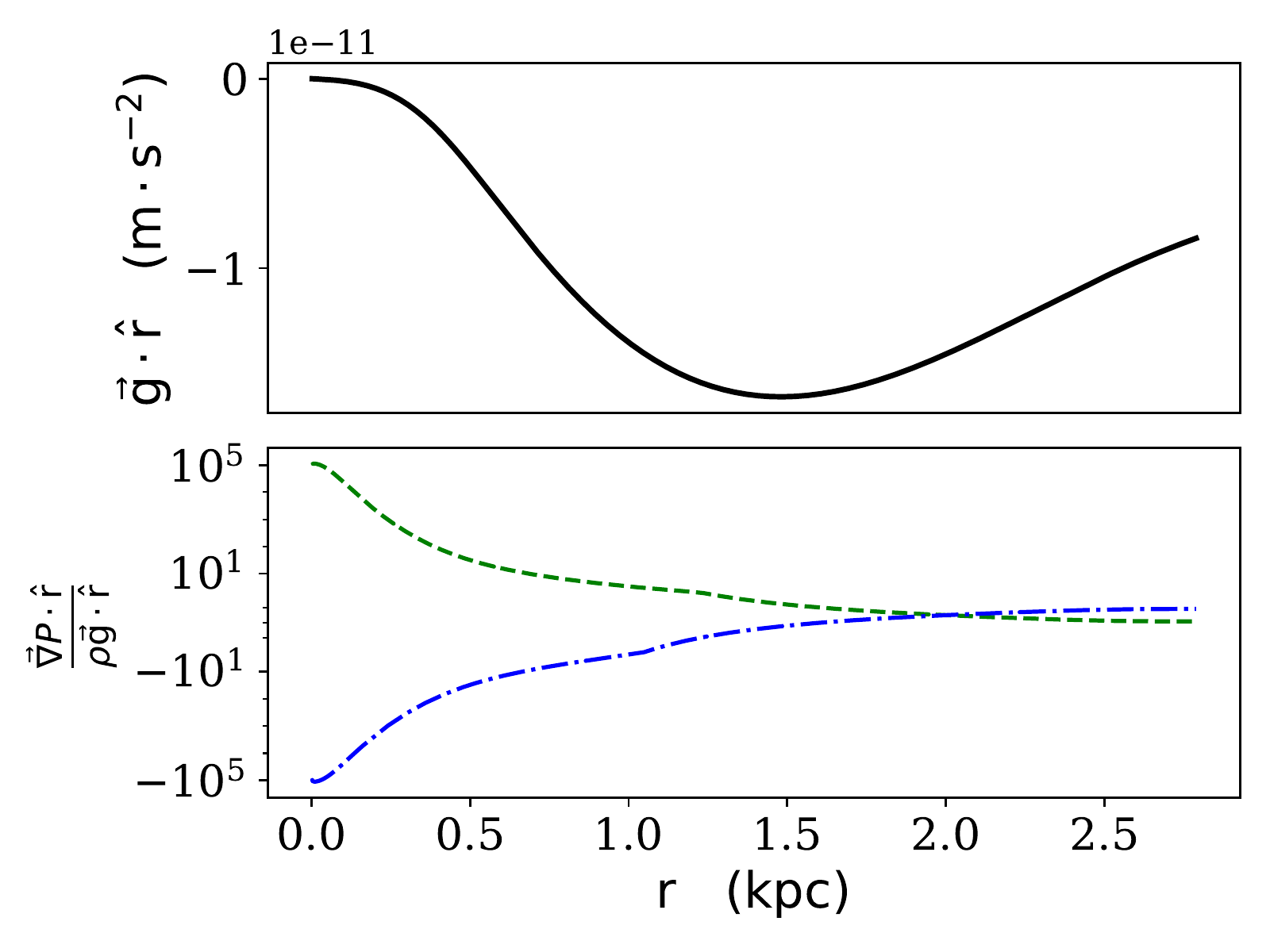}
\caption{The forces in the halos with the temperature profile of $y = \left(1 + b \xi^2\right)^{-1}$, $m=100$ eV, $\rho_0=10^{-22}$ (kg$\cdot$m$^{-3}$), and $T_0=0.1$ (Kelvin) for $b=0$ (top-left), $b=0.001$ (top-right), $b=1$ (bottom-left), and $b=10$ (bottom right).
The top panel of each plot show the radial gravitational acceleration $\vec{\mathrm{g}}\cdot \hat{\mathrm{r}} = -\partial_r\Phi$. The dashed and dashdotted curves in the bottom panel of each plot refer to the two forces of pressure divided by the gravitational force $\frac{\vec{\nabla} P_1\cdot \hat{\mathrm{r}} }{\vec{\mathrm{g}}\cdot \hat{\mathrm{r}}}$ and $\frac{\vec{\nabla} P_2\cdot \hat{\mathrm{r}} }{\vec{\mathrm{g}}\cdot \hat{\mathrm{r}}}$ respectively. 
The force of gravity is always equal to the net of the two forces of the pressure. In the central region of the non-isothermal halos, it is the pulling force of the pressure rather than gravity that maintains the stability.  \lb{fig:forces}}
\end{figure*}  
In isothermal scenarios, the high degeneracy level solution presented in figure~\ref{Fig:AllCurves2} is the most compact possible halo. However, by comparing figures \ref{Fig:AllCurves1} and \ref{Fig:AllCurves2}, we can observe that more compressed halos are possible in the presence of temperature gradient. 

Figure \ref{Fig:AllCurves1} represents some halos that are more compressed than their corresponding fully degenerate halos of figure \ref{Fig:AllCurves2} because there is a pulling force of pressure in the non-isothermal cases that is absent in the isothermal solutions.  The forces that are involved in Eq.~\eqref{Eq:HydroStat1} are the gravitational force on the right-hand side and the force due to the pressure on the left-hand side. Taking a derivative of the most general pressure in Eq.~\eqref{Eq:PressureDensity}, the force of pressure reads
\bqn
-\frac{dP}{dr} &=& -\frac{2(kT_0)^{\frac{5}{2}}}{\alpha^3}
\left(
\frac{5}{2} y^{\frac{3}{2}} f_{\frac{5}{2}}(z) \frac{dy}{dr} + 
y^{\frac{5}{2}} f_{\frac{3}{2}}(z) \frac{d\left(\Lnz\right)}{dr}
\right)\nb\\
&\equiv & -\frac{dP_1}{dr} - \frac{dP_2}{dr}.
\eqn

The first force of the pressure is proportional to $\frac{dy}{dr}$ and is absent in isothermal solutions. If the temperature gradient has a negative sign, which we expect it to have, this is a pushing force that decompresses the halo. If as in figure~\ref{Fig:PredictedTemperatureForBurkertMassProfile}, the temperature rise with the distance, the force will be inward. 

The second force of the pressure is proportional to $\frac{d\left(\Lnz\right)}{dr}$ which has a varying sign in non-isothermal solutions. It is a pulling force that compresses the halo at the center and a pushing force at the edge. In isothermal halos, it is only an outward force. 

In figure~\ref{fig:forces}, the forces are depicted for isothermal $b=0$ and non-isothermal $b=0.001$, $b=1$, and $b=10$ solutions of figure~\ref{Fig:AllCurves1}. We would like to emphasize that the first two solutions are analytically validated. In non-isothermal halos, it is the inward force of the pressure that compresses the halo at around the center, and the gravitational force is negligible. From the figure, one can observe that the strengths of the forces of the pressure are orders of magnitude higher than the strength of gravity in the presence of temperature gradient. Also note that, toward the center, $\frac{dP_2}{dr}$ has opposite signs for $b=0$ and $b \neq 0$. 

Therefore, with a temperature gradient of just 0.1 Kelvin over thousands of light-years, one can neglect the force of gravity for the first few kilo-parsecs of the halo.  With such strong additional pulling forces, it is easy to overcome the Pauli blocking forces even if DM particles are  extremely light. In \cite{2018EPJC...78..639B}, a cosmological model for such light fermions is presented that does not contradict the observations of large scale structures. Later in section~\ref{Sec:PhaseSpaceMassBounds}, we discuss that such additional attractive forces can significantly lower the so-called phase-space lower bounds on the mass of fermionic DM that is currently derived in the literature for isothermal halos in the absence of the attractive forces associated with non-isothermality. 

\begin{figure*}
\includegraphics[width=\textwidth]{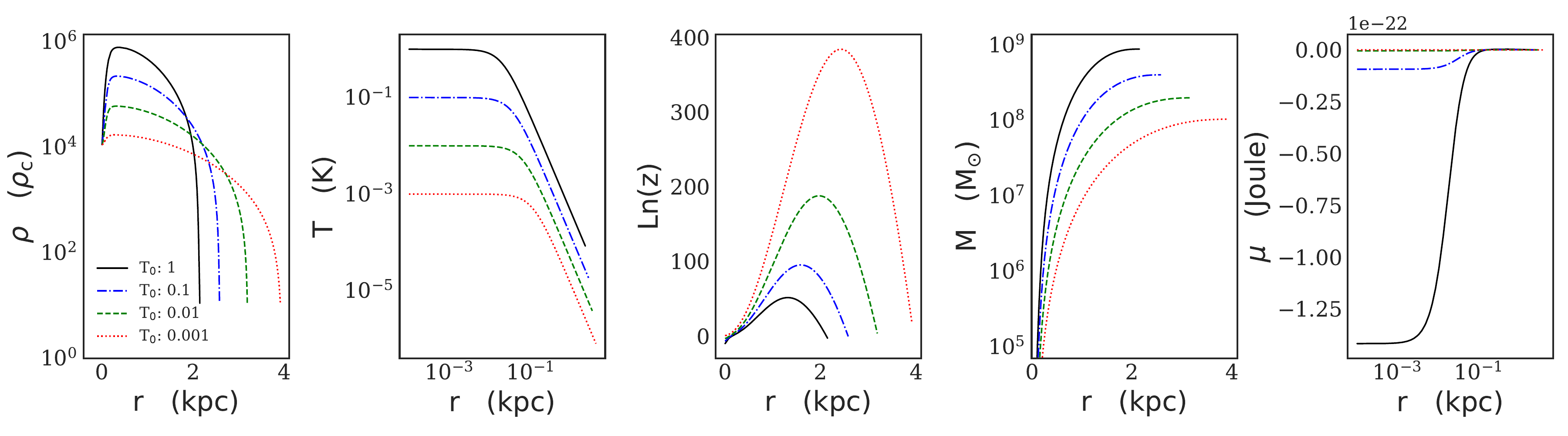}
\includegraphics[width=\textwidth]{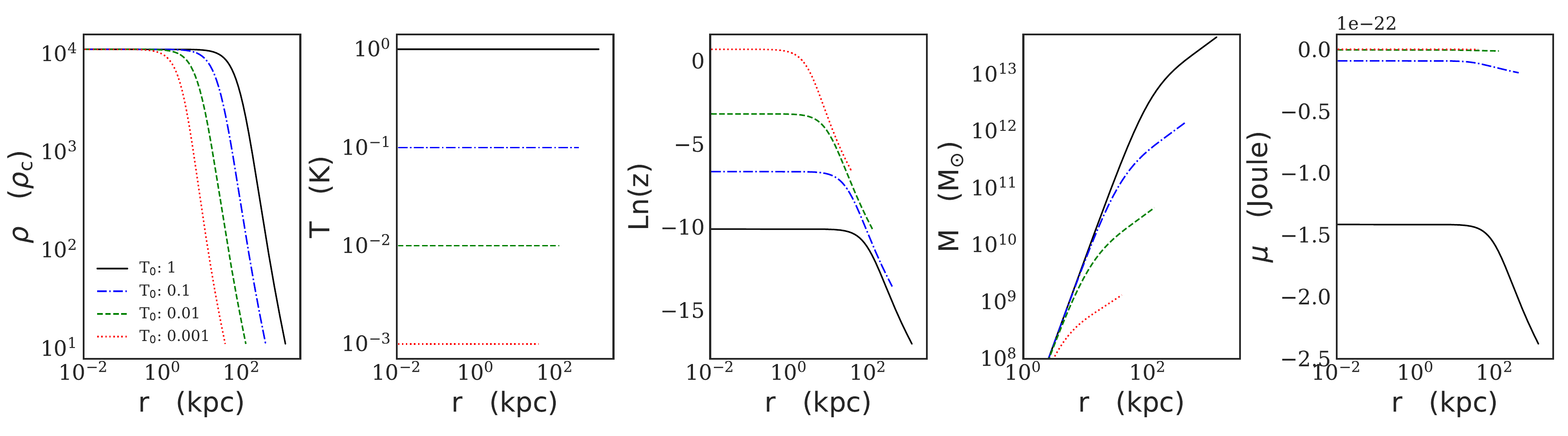}
\caption{Numerical solutions for a range different central temperatures with $m=100$ eV, and $\rho_0=10^{-22}$ (kg$\cdot$m$^{-3}$). 
(top): non-isothermal temperature profile of the form $y = \left(1 + b \xi^2\right)^{-1}$ with $b=100$. (bottom):  isothermal profiles. \lb{Fig:AllCurves3}}
\end{figure*}

The effects of increasing the central temperature are shown in figure~\ref{Fig:AllCurves3} where all of the solutions have DM mass of $100$ eV and central mass density of $\rho_0=10^{-22}$ (kg$\cdot$m$^{-3}$). Again, we use the generic profile of the form $y = \left(1 + b \xi^2\right)^{-1}$ where different $T_0$ values for both non-isothermal case of $b=100$ (top panel) and isothermal case of $b=0$ (bottom panel) are presented.  As can be seen from the figure, by increasing the central temperature, the size of halo decreases in non-isothermal solutions and increases in isothermal solutions. Interestingly, the smallest halo in figure~\ref{Fig:AllCurves3}~(top) has the same central temperature as the largest halo in figure~\ref{Fig:AllCurves3}~(bottom).

These two opposite behaviors root back to the different natures of their fugacity--or equivalently the chemical potential, profiles. When $b \neq 0$, higher $T_0$ leads to a steeper temperature gradient and therefore stronger inward force of the pressure.

\subsection{Halo encounters}
Rather than being isolated, DM halos are regularly encountering other galaxies. Such collisions are categorized depending on the sizes and velocities of the colliding galaxies. In general collisions lead to non-zero time derivative of the distribution function. 
The time evolution of the distribution function is governed by the Boltzmann equation
\bqn
\frac{df}{dt} &=& \frac{\partial f}{\partial t} + \vec{\nabla} f \cdot \vec{v} - \vec{\nabla}\Phi \cdot \frac{\partial f}{\partial \vec{v}} \nb\\
&=& C[f],
\eqn
where $C[f]$ is a functional of the distribution function accounting for the collision, i.e., the gravitational forces due to the other system, $\Phi$ is the self-gravitational potential, and $\vec{v}$ is the velocity of DM particles.

When the encounter is long term, as in the case of major mergers, such that the time derivative of the distribution function is non-perturbatively different from zero, the system is not stable and stability solution as the subject of this paper becomes irrelevant.

When galactic collisions are fast, the distribution function deviates slightly from the initial form. Therefore, an approach similar to the one employed for studying the anisotropies in the cosmic microwave background should be viable. In this case, the distribution function of fermionic DM can be imagined to have a form similar to the Fermi-Dirac distribution
\bqn
f &=& \frac{1}{\big(z+\delta z\big)\exp\Big(\big(\beta+\delta \beta \big)\varepsilon \Big)+1},
\eqn
where we assume that $\delta z \ll z$ and $\delta \beta \ll \beta$. After expanding around the initial Fermi-Dirac distribution and neglecting the second order terms, the distribution function reads
\bqn
&\simeq & \frac{1}{z e^{\beta \varepsilon}+1} 
- \frac{e^{\beta \varepsilon} \Big(\varepsilon z \delta \beta +\delta z\Big)}{\Big(z e^{\beta
   \varepsilon}+1\Big)^2}
+{\cal{O}}\left(\delta^2\right).
\eqn

During the halo encounter, the collision term $C[f]$  can be found from for example \cite{2008gady.book.....B} using the cross-section of the gravitational interactions. 
After the encounter, the partial derivative of $f$ with respect to time is still non-zero.  
The stability Eq.~\eqref{Eq:HydroStat1} is only valid at the very end of the process when $\frac{\partial f}{\partial t}=0$. However, since both sides of the Boltzmann equation can be expanded in terms of $\delta \beta$ and $\delta z$, and since the zeroth and first order equations are independent, the stability solutions presented in this paper are always valid to the zeroth order. 

In summary, the stability solutions of this paper are only applicable to halos that have not been involved in a major merger in the past few Giga-years. Recent studies show that the probability of such encounters is higher for more massive halos \cite{2015MNRAS.450.1604L, 2017MNRAS.464.1659Q, 2020arXiv200102687O}. 

One interesting halo encounter is when the transferred energy is entirely spent to overcome the gravitation bound of the outer regions and leads to mass loss, such that $C[f]$ and $\frac{\partial f}{\partial t}$ are only non-zero in the outer regions of the halo. 
Therefore, the stability solutions presented in this paper are still valid in the interior regions because the gravitational force at any distance is only a function of the enclosed mass and the balance of the forces in the inner regions will not be changed significantly.  

Tidally truncated subhalos submerged in a host are in general subject to at least dynamical friction which can significantly change their temperature profile and hence the density profile. However, due to Gauss's law, the gravitational force of the interior region is not affected by the mass loss and the stability solutions presented in this paper are well justified up until the limiting radius provided that the temperature profile reflects the encounter.

\section{Phase-space mass bounds on DM mass}
\lb{Sec:PhaseSpaceMassBounds}

In 1979, Tremaine and Gunn derived the first lower limit on the mass of DM \cite{1979PhRvL..42..407T}. The derivation depends on a set of assumptions whose validities are not known yet. More specifically they assumed (I) a specific primordial phase-space density, (II) DM is collision-less, i.e., the maximum of its phase-space density is conserved, (III) galactic DM has a Maxwell-Boltzmann distribution, and (IV) DM halo is isothermal. The first two assumptions are not valid for interacting DM. The last two assumptions are also not valid in the degenerate non-isothermal models in which we are interested. 
The Tremaine-Gunn bound is more related to a knowledge of the primordial phase-space and its evolution over time (which are model dependent) than the Fermi-Dirac statistics of particles. The same bounds apply to some non-fermionic models of dark matter \cite{2001ApJ...561...35D}. 

\begin{figure*}
\includegraphics[width=\linewidth]{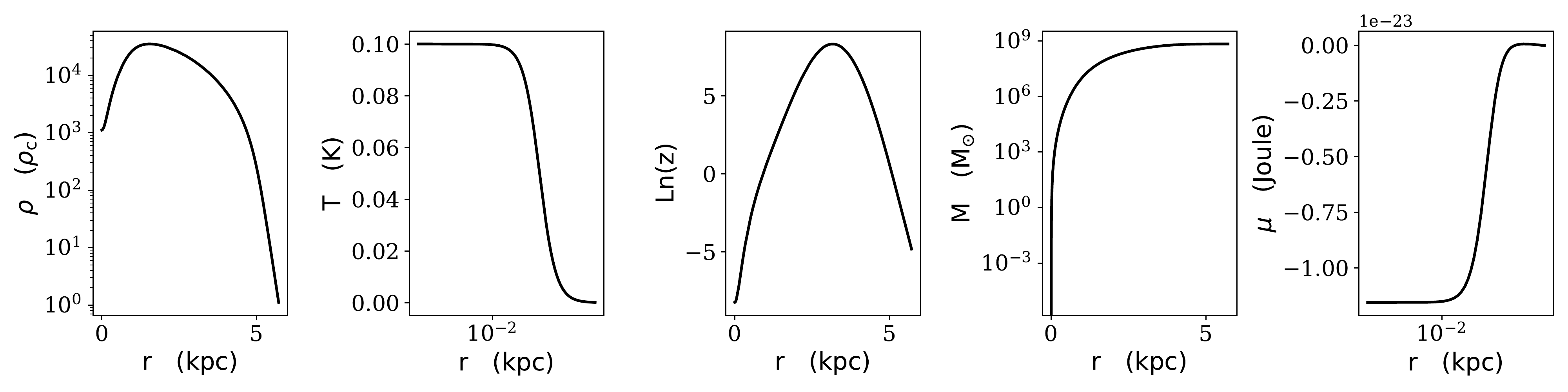}
\caption{Numerical solutions for non-isothermal temperature profile of the form $y = \left(1 + 10 \xi^2\right)^{-1}$, $m=75$ eV, and $T_0=0.1$ (Kelvin). \lb{Fig:CurvesOfm75}}
\end{figure*}
True models of galactic fermionic DM were later studied in for instance \cite{2001astro.ph.11366B, 2002PhRvE..65e6123C, 2014MNRAS.442.2717D}. 
A lower bound on the mass of a genuinely fermionic DM can be derived using the lower limit of its dispersion velocity at the full degeneracy level, see for instance \cite{2009JCAP...03..005B, 2018MNRAS.475.5385D}
\bqn
\lb{Eq:dispersionInequlity}
\sigma_{\text{F.D}}^2 \geq \sigma_{\text{full-deg.}}^2.
\eqn
If the Maxwell-Boltzmann distribution governs a coarse-grained fermionic DM, its dispersion velocity still needs to be larger than the minimum in Eq.~\eqref{Eq:dispersionInequlity}, and the inequality leads to a lower limit on the mass of DM.  
In \cite{2015JCAP...01..002D}, it is discussed that the inequality is trivial if DM halo is made of a degenerate Fermi gas. As is mentioned by \cite{2018MNRAS.475.5385D}, the inequality is trivial as far as the Fermi-Dirac distribution is used to describe DM halo regardless of its degeneracy level, i.e. even if $z \ll 1$. To see this, we start with the definition of the dispersion velocity
\bqn
\sigma^2 \equiv \frac{kT}{m}h(z),
\eqn
whose minimum is equal to 
$\frac{1}{5}\left( \frac{3\rho \pi^2 \hbar^3}{m^4}\right)^{\frac{2}{3}}$. Inserting this and the exact equation for the dispersion velocity into Eq.~\eqref{Eq:dispersionInequlity} and a straightforward calculation leads to 
\bqn
f_{\frac{5}{2}}(z) \geq \frac{6^{\frac{2}{3}}\pi^{\frac{1}{3}}}{10}\left( f_{\frac{3}{2}}(z)\right)^{\frac{5}{3}}.
\eqn
Even in the classical limit where the Fermi-Dirac distribution is effectively Maxwell-Boltzmann, the inequality is trivial. This can be seen by replacing the Fermi-Dirac integrals with their low-fugacity approximation. For this reason, our software exclusively (even in effectively Maxwellian DM halos) works with a Fermi-Dirac distribution such that the limitation of the phase-space is always respected. 

Also, as is mentioned in \cite{2018MNRAS.475.5385D}, if DM distribution is not entirely Maxwellian, it is not possible to use the dispersion velocity of entirely Maxwellian visible matter to learn the escape velocity of DM. Such learning becomes even less possible in non-isothermal halos because the dispersion velocity is a function of temperature profile and the mechanisms of heating the visible matter are due to the release of the stored potential energy of electric and strong forces, as well as gravity-based mechanisms. On the contrary, DM halos are most likely heated up by gravity-based mechanisms and friction. As can be seen from figure~\ref{Fig:escapeVelocity}, in the outer regions of non-isothermal halos, the kinetic energy of DM particles are low enough that they cannot escape the gravitational well.

However, in \cite{2018MNRAS.475.5385D}, the lower bound of $\sim 100$~eV for the mass of DM is derived if the observed dwarf galaxies are infinitely degenerate.   
The basis for the bound is that the decay period of satellite galaxies due to the Chandrasekhar friction has to be larger than $10^{10}$ years. 
Chandrasekhar's estimation for the decay time is a function of (i) the velocity of the satellite, (ii) its distance from the center of the host galaxy, and (iii) mass and radius of the halo of the satellite.  

Among the three enumerated factors above, only the mass and radius of a satellite's halo depend on the stability Eq.~\eqref{Eq:HydroStat_Dimensionless}. If two solutions to the stability equation have the same halo mass and radius, their corresponding decay time will be the same.
The lower bound of $\sim 100$~eV in the reference above is derived for entirely degenerate halos--as the most compact possible solutions of isothermal scenarios. However, there is no observation confirming that such halos are infinitely degenerate. Due to the stronger frictional forces that such galaxies experience, it is likely that the halos are non-isothermal in which case the same halo size and mass are possible with lower DM mass.  
For example, 
in figure~\ref{Fig:CurvesOfm75}, we present a non-isothermal halo made of DM mass of 75 eV that has similar halo mass and radius as in its corresponding fully degenerate solution presented in figure~\ref{Fig:AllCurves2} and made with DM mass of $100$ eV.

\begin{figure}
\centering
\includegraphics[width=0.49\columnwidth]{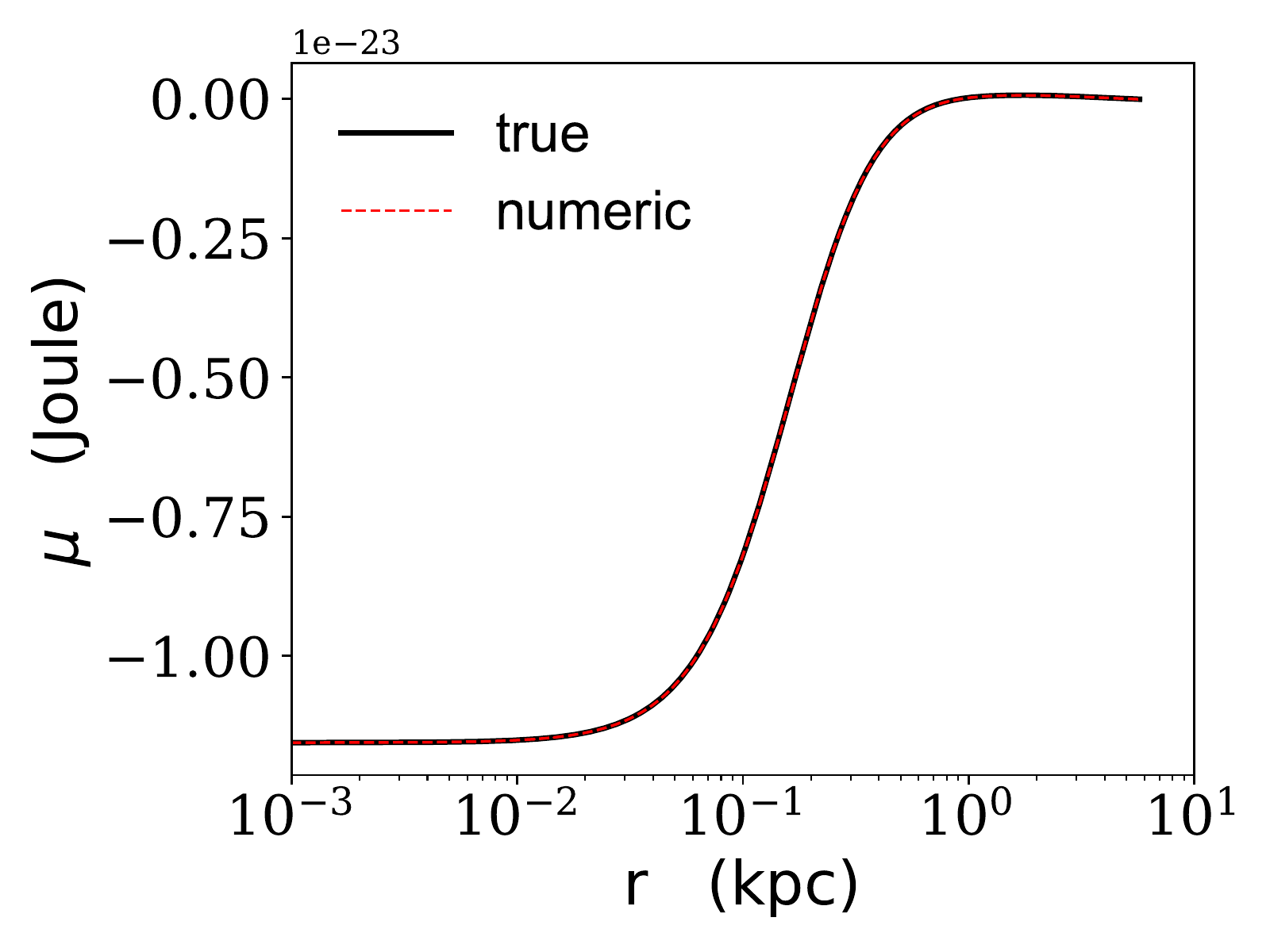}
\includegraphics[width=0.49\columnwidth]{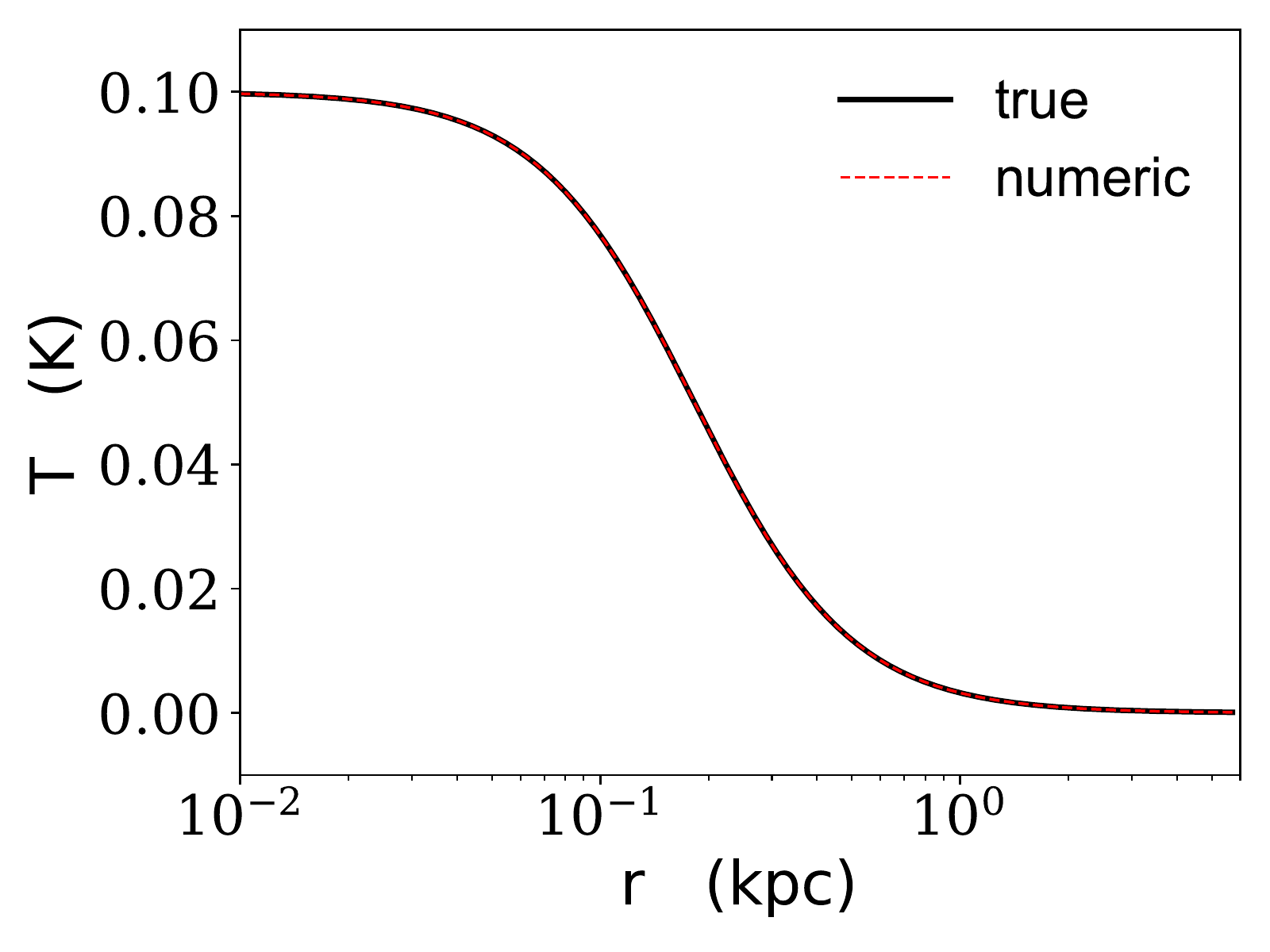}
\caption{The chemical potential and temperature of figure~\ref{Fig:CurvesOfm75}--labeled true, are reproduced--labeled numeric, by numerically solving Eq.~\eqref{Eq:IntegralEquation} and using Eq.~\eqref{Eq:Temperature_inTermsOf_rho_z} after assuming that the mass density in figure~\ref{Fig:CurvesOfm75} with $T_0=0.1$~(K) is the true function. \lb{Fig:Mu_and_T_m75} 
}
\end{figure}

At this point, we prove that if the temperature is not modeled and left as a free parameter, it is not possible to place a lower bound on DM mass. 
We start by rephrasing the mass density in Eq.~\eqref{Eq:PressureDensity} as
\bqn
\lb{Eq:Temperature_inTermsOf_rho_z}
T = \frac{\alpha^2}{\left(2m\right)^{\frac{2}{3}}k}\left(\frac{\rho}{f_{\frac{3}{2}}(z)}\right)^{\frac{2}{3}},
\eqn
and note that the temperature cannot take negative values in this equation. We use this equation to eliminate the temperature in the Pressure in Eq.~\eqref{Eq:PressureDensity}
\bqn
P = \frac{2\alpha^2}{\left(2m\right)^{\frac{5}{3}}}\frac{f_{\frac{5}{2}}(z)}{\left(f_{\frac{3}{2}}(z)\right)^{\frac{5}{3}}}\rho^{\frac{5}{3}}. 
\eqn

At this point, we use observations to define a mass profile, such that the mass density and its derivative are known functions of $r$. In doing so, we also pay special attention that $\rho(r)$ is set such that the decay period due to Chandrasekhar friction is consistent with expectations. The derivative of the pressure with respect to $r$ reads
\bqn
\frac{dP}{dr} = \frac{\partial P}{\partial z}\frac{dz}{dr}+\frac{\partial P}{\partial \rho}\frac{d\rho}{dr} \equiv A(r,z)\frac{d\Lnz}{dr}+B(r,z),
\eqn
where $A(r,z)=\frac{\partial P}{\partial z}z$, and $B$ is the last term and are given by
\bqn
&&A(r,z) = \frac{2\alpha^2}{\left(2m\right)^{\frac{5}{3}}}\rho^{\frac{5}{3}}
\left(
\frac{1}{\left(f_{\frac{3}{2}}(z)\right)^{\frac{2}{3}}}-
\frac{5}{3}\frac{f_{\frac{1}{2}}(z)f_{\frac{5}{2}}(z)}{\left(f_{\frac{3}{2}}(z)\right)^{\frac{8}{3}}}
\right),\nb\\
&&B(r,z) = \frac{5}{3}\frac{2\alpha^2}{\left(2m\right)^{\frac{5}{3}}}\rho^{\frac{2}{3}}\frac{d\rho}{dr}\frac{f_{\frac{5}{2}}(z)}{\left(f_{\frac{3}{2}}(z)\right)^{\frac{5}{3}}}.
\eqn

Inserting equations above into the stability Eq.~\eqref{Eq:HydroStat1}, we will arrive at a non-linear first-order differential equation for the degeneracy of DM halo
\bqn
\frac{d\Lnz}{dr} = \frac{-B(r,z)-G\rho M(r)/r^2}{A(r,z)},
\eqn
which satisfies the initial condition at $r=0$ if $\rho$ is set such that its derivative at the center is zero. It is important to note that since $\rho(r)$ is now a known function of $r$, the only unknown parameter is the fugacity $z$.
After an integration of the differential equation, we arrive at
\bqn
\lb{Eq:IntegralEquation}
\text{Ln}\left(\frac{z}{z_0}\right) = -\int_0^r  \frac{B(r',z)+G\rho M(r')/r^{'2}}{A(r',z)} dr'.
\eqn

Although this is a complex integral equation for the degeneracy level of DM halo, 
it eventually has a solution for any arbitrarily light DM mass $m$. Since the degeneracy level of DM halo is not observable at this time, the only way to constrain the solutions to this integral equation is through building temperature profiles for DM halos. In doing so, we should note that the temperature profiles are not universal, see section~\ref{Sec:TempGrad}. The temperature profiles of satellite galaxies are especially different than the temperatures of large galaxies due to higher friction, higher contraction, and the tidal effects of the host galaxy.

In figure \ref{Fig:Mu_and_T_m75}, we present the numerical solution of the integral Eq.~\eqref{Eq:IntegralEquation} by feeding to it the mass density corresponding to $T_0=0.1$ in figure \ref{Fig:CurvesOfm75} and then using Eq.~\eqref{Eq:Temperature_inTermsOf_rho_z}. Since two independent numerical methods lead to the same set of solutions, we can take this as another validation on top of the interval reduction validation method discussed in section \ref{Sec:SoftWareValidation}.  

It is important to note that the escape velocity and the phase-space limitation arguments mentioned above are trivially satisfied by the solution to Eq.~\eqref{Eq:IntegralEquation}. Because, this is a stability equation whose solution guarantees that DM particles are trapped in the halo. Also, since the full Fermi-Dirac EOS is used for the derivation, the limitation of phase-space is naturally met. 

Finally, we would like to emphasize that visible matter is often denser than DM at distances close to the center of halos. The extra gravitational force helps to contract DM even further. Since the gravitational force due to visible matter can be significant, fermionic DM halos are compacted much more than if the visible matter is neglected. Therefore, even in isothermal halos, the lower bound on the mass of DM is not as low as is currently derived.

\section{Conclusion}
\lb{Sec:conclusion}
We have studied non-isothermal non-interacting fermionic spherical dark matter halos. Using the full EOS of Fermi-Dirac statistical systems, we derive the most general stability equation and present computer software to numerically solve it. Since the full Fermi-Dirac EOS is used in the software, the transitions between degeneracy levels are smooth and the limitation of fermionic phase-space is never violated in the numerical solutions. From non-degenerate to highly degenerate Fermi halos with any temperature profile can be investigated with the software.

We have studied non-isothermal halos using a generic temperature profile of the form $T = T_0\left(1+\left(\frac{r}{r_0}\right)^2\right)^{-1}$, and shown that their chemical potential profile is substantially different from that of the corresponding isothermal halos. We show that the mass and radius of such non-isothermal halos decrease by increasing the temperature gradient. We have shown that the force due to the pressure has inward as well as outward components, and at the central regions of the studied halos, it is the inward force of the pressure, rather than gravity, that maintains the stability. 

We have discussed the phase-space lower bounds on the mass of DM as well as the importance of modeling the temperature profile of DM halos for deriving them. We have shown that if the temperature is left as a free parameter, any arbitrarily light DM mass can explain the observed mass profile of DM halos. 

It has been discussed that the limitation of the phase-space of fermions does not restrict their configuration volume.
It is the inward force acting on the fermions that determines the size of DM halos. In the presence of temperature gradient, the inward force due to the pressure adds to the inward force of gravity and maintains the stability of compressed fermionic halos. We have shown examples where the former force is orders of magnitude stronger than the latter for a rather vast region. The inward component of the force of pressure is absent in isothermal halos.

We have shown that if the quantum nature of DM is irrelevant in the halos, the temperature profile is analytically given in terms of the mass profile. By requiring that the temperature is not negative, we place an upper bound on the mass of DM. We find that if the central temperature of DM halo is only a few Kelvins, the mass of DM cannot be larger than a few keV.

\section*{Acknowledgement}
We would like to thank the anonymous reviewers as well as the editors of EPJC for their useful suggestions that improved the reported work.


\bibliographystyle{elsarticle-num-names} 
\bibliography{Refs}

\end{document}